\documentclass[a4paper,12pt]{article}
\pdfoutput=1
\usepackage{epsfig,graphicx,xcolor,amsbsy,amssymb,latexsym,amsfonts,amsmath,,tcolorbox,setspace}
\usepackage{pstricks}
\usepackage{color}
\usepackage{soul}
\usepackage{placeins}
\usepackage{wrapfig,framed,caption}
\usepackage[makeroom]{cancel}
\usepackage{todonotes}
\usepackage{stmaryrd}

\usepackage{wasysym}

\usepackage{multirow}

\definecolor{shadecolor}{gray}{0.925}

\usepackage{eurosym}
\usepackage[a4paper]{geometry}
\usepackage{hyperref}
\usepackage{graphicx}
\usepackage{tikz}
\usetikzlibrary{decorations.pathreplacing,decorations.markings,snakes}
\usetikzlibrary{decorations.pathmorphing}
\usetikzlibrary{patterns}
\usetikzlibrary{calc}
\usepackage{xcolor}
\usepackage{amsmath,amssymb,amsfonts,pstricks,setspace}
\usepackage[enableskew]{youngtab}

\tikzset{->-/.style={decoration={
  markings,
  mark=at position .55 with {\arrow{stealth}}},postaction={decorate}}}

\tikzset{-<-/.style={decoration={
  markings,
  mark=at position .55 with {\arrow[>=stealth]{<}}},postaction={decorate}}}  

\usepackage[style=numeric-comp,sorting=none,maxbibnames=9,backend=bibtex]{biblatex}
\numberwithin{equation}{section}

\usepackage{wrapfig}

\newcommand{\bea}{\begin{eqnarray}\displaystyle}
\newcommand{\eea}{\end{eqnarray}}

\newcommand{\ser}{\mathcal{S}}

\newcommand{\ztu}[1]{\mathfrak{z}(#1)}
\newcommand{\ztua}[2]{\mathfrak{z}_+(#1,#2)}
\newcommand{\ztub}[2]{\mathfrak{z}_-(#1,#2)}

\newcommand{\dop}{\mathfrak{d}}

\newcommand{\madj}{\mathfrak{m}}

\newenvironment{linesmall}[1]
  {\arraycolsep=3pt\scriptsize
   \array{#1}}
  {\endarray}

\title{
{\bf $\mathcal{N}=2^*$ Partition Function on $\mathbb{S}^4$ and\\[6pt] Gauss-deformed Zeta functions}\\[30pt]}

\author{\large \textsc{Timothé Alezraa\footnote{\tt t.alezraa@ip2i.in2p3.fr}}\,,\hspace{0.4cm}\textsc{Stefan~Hohenegger\footnote{\tt s.hohenegger@ip2i.in2p3.fr}}\,,\hspace{0.3cm}and\hspace{0.2cm} \textsc{Taro Kimura\footnote{\tt taro.kimura@ube.fr}} }

\begin{document}

\maketitle
\thispagestyle{empty}

\maketitle
\thispagestyle{empty}
${}$\\
\begin{center}
\renewcommand{\thefootnote}{\fnsymbol{footnote}}\vspace{-0.5cm}
${}^{\footnotemark[1]\footnotemark[2]}$ Universit\'e Lyon 1, CNRS, IP2I, UMR 5822, Villeurbanne, France\\[0.5cm] 
${}^{\footnotemark[3]}$ Universit\'e Bourgogne Europe, CNRS, IMB UMR 5584, 21000 Dijon, France\\ Institut Universitaire de France (IUF), France\\[2.5cm]
\end{center}

\begin{abstract}
Using localisation techniques, the partition function of $\mathcal{N}=2^*$ Super-Yang-Mills theory on $\mathbb{S}^4$ with gauge group $U(2)$, can be computed explicitly in integral form. At any fixed order in the mass expansion, its contributions take the form of asymptotic series in the gauge coupling, with vanishing radius of convergence. Resurgence techniques can be further used to provide physical meaning to these expansions, however, the resulting expressions generally remain complicated (albeit convergent) series. In this work, focusing on the perturbative contribution to the partition function, we propose to alternatively regularise the asymptotic series, which allows us to express them as finite combinations of \emph{Gauss-deformed zeta functions}. These functions resemble Mellin representations of the Riemann zeta function supplemented by Gaussian kernels. Further organising the mass expansion according to the \emph{depth} of these functions, allows us to express the leading terms in a remarkably compact form. Moreover, this organisation reveals non-trivial relations among specific contributions of the perturbative partition function.
\end{abstract}

\newpage

\tableofcontents

\section{Introduction}

Gauge theories with extended supersymmetry constitute a concrete arena to study exact results at finite coupling. Notably the study of $\mathcal N=2$ gauge theories in four dimensions (as well as their higher dimensional counterparts) has a rich history, with many explicit results related to perturbative and non-perturbative observables. 
A class of examples, which has received a lot of attention in the literature, is the partition function of $\mathcal{N}=2$ theories on $\mathbb{S}^4$ \cite{Pestun:2007rz} (and more generally on ellipsoidal deformations \cite{Hama:2012bg}). This background preserves a supercharge, whose square generates (among other transformations) rotations on $\mathbb{S}^4$, and which can be used to compute the partition function via localisation in a very explicit form. More concretely, the path integral of $\mathcal{N}=2$ theories with gauge group $G$ can be reduced to a finite-dimensional integral over the Coulomb moduli, taking a value in the Cartan subalgebra $a \in \mathfrak{t} \subset \operatorname{Lie}(G)$, which can schematically (\emph{i.e.} up to normalisation factors) be written in the following form
\begin{equation}
 \mathcal{Z}_{}(\omega)=\int_{\mathfrak t}[da]\; Z_{\mathrm{cl}}(a;\omega)\,Z_{\mathrm{1\text{-}loop}}(a;\omega)\,Z_{\mathrm{inst}}^{N}(a;\omega)\,Z_{\mathrm{inst}}^{S}(a;\omega),
 \label{eq:intro_sphere_localization}
\end{equation}
Here $Z_{\mathrm{cl}}$ is a Gaussian factor (stemming from the classical part in the original path integral), while $Z_{\mathrm{1\text{-}loop}}$ represents the 1-loop determinant together with an integral measure denoted by $[da]$. Finally, $Z_{\mathrm{inst}}^{N}$ and $Z_{\mathrm{inst}}^{S}$ 
denote the instanton partition functions 
\cite{Losev:1997tp,Moore:1997dj,Lossev:1997bz,Nekrasov:2002qd} associated with the local 
$\mathbb{R}^4$ patches around the north and south poles 
of $\mathbb{S}^4$, respectively. 
Since they are conjugate to each other, $\overline{Z_{\mathrm{inst}}^{N}} = Z_{\mathrm{inst}}^{S}$, we simply write the instanton part $|Z_{\mathrm{inst}}(a;\omega)|^2 = Z_{\mathrm{inst}}^{N}(a;\omega) Z_{\mathrm{inst}}^{S}(a;\omega)$.
Furthermore, $\omega$ collectively denotes various parameters of the theory, 
including mass parameters associated with flavour symmetries, gauge couplings, theta angles, and deformation parameters entering the partition 
functions in the case of an ellipsoid. 
A unified perspective on these parameters is provided by 
brane constructions in string theory, in which they admit 
a natural geometrical interpretation in terms of the 
underlying ten-dimensional spacetime.

In this paper, we are exclusively interested in the $\mathcal{N}=2^*$ Super-Yang-Mills (SYM) theory obtained by giving a non-zero mass $\madj$ to the adjoint hypermultiplet of the four-dimensional $\mathcal N=2$ theory. We shall focus on the gauge group $G=U(2)$ with the self-dual $\Omega$-background $\varepsilon_1=-\varepsilon_2=1$ and the gauge coupling $g$.
Moreover, we study in detail the mass expansion of the perturbative partition function (see \eqref{eq:Z0} for the definition)%
\footnote{The study of the mass dependence of the sphere partition function is also motivated by applications to integrated correlators in $\mathcal{N}=4$ theories (see \emph{e.g.} \cite{Pestun:2007rz,Gomis_2012,dorigoni2024electromagneticdualitylinedefect}). Indeed, certain integrated four-point functions can be extracted from derivatives of the $\mathcal{N}=2^*$ sphere free energy with respect to the adjoint mass and the gauge coupling, evaluated at the maximally supersymmetric point. Their exact dependence on the complexified coupling, and link with modularity and electromagnetic duality is explored in \cite{Dorigoni:2021guq}. } 
\begin{align}
&\mathcal{Z}_{\emptyset}(g,\madj)=\sum_{s=0}^\infty \madj^{2s}\,\mathcal{Z}_\emptyset^{(2s)}\,.
\end{align}
The terms $\mathcal{Z}_\emptyset^{(2s)}$ are integrals over the real line $\mathbb{R}$, which are rendered convergent due to a Gaussian factor stemming from $Z_{\text{cl}}$. Exploiting this Gaussian factor, allows to explicitly perform the integral using the Laplace method (see \emph{e.g.} \cite{BenderOrszag} for a review). In order to further discuss the result, we find it useful to further decompose the contribution at order $\mathcal{O}(\madj^{2s})$ into finitely many further terms
\begin{align}
&\mathcal{Z}_\emptyset^{(2s)}=\sum_{r=0}^s \mathcal{Z}_\emptyset^{(2s),r}\,,\label{InitialMassExpansion}
\end{align}
where $\mathcal{Z}_\emptyset^{(2s),r}$ is an infinite series whose general term is made up from products of exactly $r$ Riemann zeta functions $\zeta_n$ (see (\ref{DefZetaFunction}) for the definition for $n>1$). Using the nomenclature from the literature on multi-zeta values, we shall therefore refer to $r$ as the \emph{depth} of the decomposition. More correctly, the $\mathcal{Z}_\emptyset^{(2s),r}$ are \emph{asymptotic} series (see \cite{Dorigoni:2021guq,Aniceto:2026pcz}) in powers of the gauge coupling $g$, \emph{i.e.} they have zero radius of convergence and are therefore only formal expressions. In order to still give a meaningful sense to these expressions, we follow two complementary strategies.
\begin{itemize}
\item[\emph{(i)}] {\bf Borel resummation:} Resurgence and resummation techniques have been successfully in many similar context (see Appendix~\ref{App:ResurgenceReview} for a brief mathematical review): for example, the perturbative expansions of certain three- and four-dimensional theories calculated via localisation, leads to Borel-summable weak-coupling series~\cite{Russo:2012kj}, which can further be analysed using resurgent techniques~\cite{Aniceto:2014hoa}. Resurgence has also been used to investigate large-$N$ expansions integrated correlators in $\mathcal{N}=4$ theories \cite{Dorigoni:2024dhy} (also in the presence of half-BPS line defects \cite{dorigoni2024electromagneticdualitylinedefect}). In the current context, the $\mathcal{Z}_\emptyset^{(2s),r}$ are of Gevrey-1 type and their Borel transformation only has singularities for negative real part in the Borel plane. This allows to resum the $\mathcal{Z}_\emptyset^{(2s),r}$ in a unique fashion, without the appearance of a Stokes phenomenon (see \cite{Aniceto:2014hoa}). In fact, we provide evidence up to $r=3$ that the $\mathcal{Z}_\emptyset^{(2s),r}$ can be written as convergent sums of a single building block $\ser$ (which is reviewed in Appendix~\ref{App:BorelResumSer}), acted upon by differential operators with respect to the coupling constant $g$. Borel resumming $\ser$ allows to write $\mathcal{Z}_\emptyset^{(2s),r}$ in the form of ($r$ nested) sums of error-functions of the coupling constant and their derivatives. The appearance of the latter has previously been observed in \cite{Aniceto:2026pcz} in a related context.
\item[\emph{(ii)}] {\bf Regularisation:} the intrinsic connection of the depth $r$ contribution $\mathcal{Z}_\emptyset^{(2s),r}$ with Riemann zeta functions affords an alternative way of making sense of them: in a first step, we replace any of the zeta functions by their Mellin representation (see (\ref{FormZetaIntegSinh})) as an integral over $\mathbb{R}_+$. In order to furthermore pull the summation under the integrals we regularise the latter at infinity with a cut-off $1/\delta$ (with $\delta\ll 1$). The resulting summation in fact converges (uniformely) leading to a finite contribution in the limit $\delta\to 0$. To leading order (\emph{i.e.} $s=1$) we obtain the same result as in \cite{Dorigoni:2021guq}, while for generic $s$ (and up to $r=3$) we show that we can write the result as finite combinations of a class of functions of the coupling $g$ that we call \emph{Gauss-deformed zeta functions}. Indeed, in their simplest incarnation, $\mathfrak{z}(a)$ (with $a\in\mathbb{R}$), they are simply Mellin-representations of Riemann zeta functions with a ($g$-dependent) Gaussian kernel. We argue that the depth 1 contribution for generic $s$ (\emph{i.e.} $\mathcal{Z}_\emptyset^{(2s),1}$) can be completely written in terms of $\mathfrak{z}(2s-1)$, acted upon by a simple differential operator in $g$. We furthermore argue (by analysing contributions up to $r=3$) that higher depth contributions contain generalisations $\mathfrak{z}(a_1,\ldots,a_n)$ that can be written as $n$ Mellin integrals with different insertions of Gauss kernels. We conjecture a set of basis functions suitable for any depth $r$.
\end{itemize}

\begin{figure}[h]
\begin{center}
\begin{tikzpicture}
\node at (0,0) {\fbox{\parbox{7.8cm}{\centering {\color{red}{asymptotic series}} {\footnotesize (\ref{SeriesZetaSingle})}\\[6pt]
$-\frac{16\pi^2}{\Gamma(2s+1)}\sum_{k=1}^\infty (-1)^k\frac{\Gamma(2k+2s-1)g^{2k-2}}{\Gamma(k)\,(16\pi^2)^k}\,\zeta_{2k+2s-1}$}}};
\draw[ultra thick,->] (2,-1) -- (4,-3);
\draw[ultra thick,->] (-2,-1) -- (-4,-3);
\node[rotate=45] at (-3.3,-1.85) {\footnotesize Borel resummation};
\node[rotate=-45] at (3.25,-1.8) {\footnotesize regularisation};
\node at (-4.8,-4.2) {\fbox{\parbox{7cm}{\centering {\color{blue}{sum of error-functions}} {\footnotesize (\ref{SeriesBorelResumCorr})}\\[6pt]
$\frac{8\pi^2\left(\dop\right)_{2s-1}}{g^2(2s)!}\sum_{\ell=1}^\infty\frac{\frac{2\,\pi^{\frac{3}{2}}\,\ell}{g}\,e^{\frac{4\pi^2\ell^2}{g^2}}\,\text{Erfc}\left(\tfrac{2\pi\ell}{g}\right)+1}{\ell^{2s-1}}$}}};
\node at (4.8,-4.2) {\fbox{\parbox{7cm}{\centering {\color{green!50!black}{Gauss-deformed zeta function}} {\footnotesize (\ref{SingleZetReg})}\\[6pt]
$\mathfrak{z}(2s+1)=\frac{1}{\Gamma(2s+1)}\int_0^\infty\frac{dw\,w^{2s}}{e^w-1}\,e^{-\frac{w^2 g^2}{16\pi^2}}$}}};
\end{tikzpicture}
\caption{Schematic relation between an asymptotic series (with vanishing radius of convergence), a convergent sum over error-functions (with $\text{Erfc}(z)=1-\text{Erf}(z)$ and $(z)_{n}$ denoting the Pochhammer symbol~\eqref{DefPochhammer}, while $\dop=g\tfrac{d}{dg}$) and the newly introduced Gauss-deformed zeta functions.}
\label{Fig:RelationAsymSeries}
\end{center}
\end{figure}
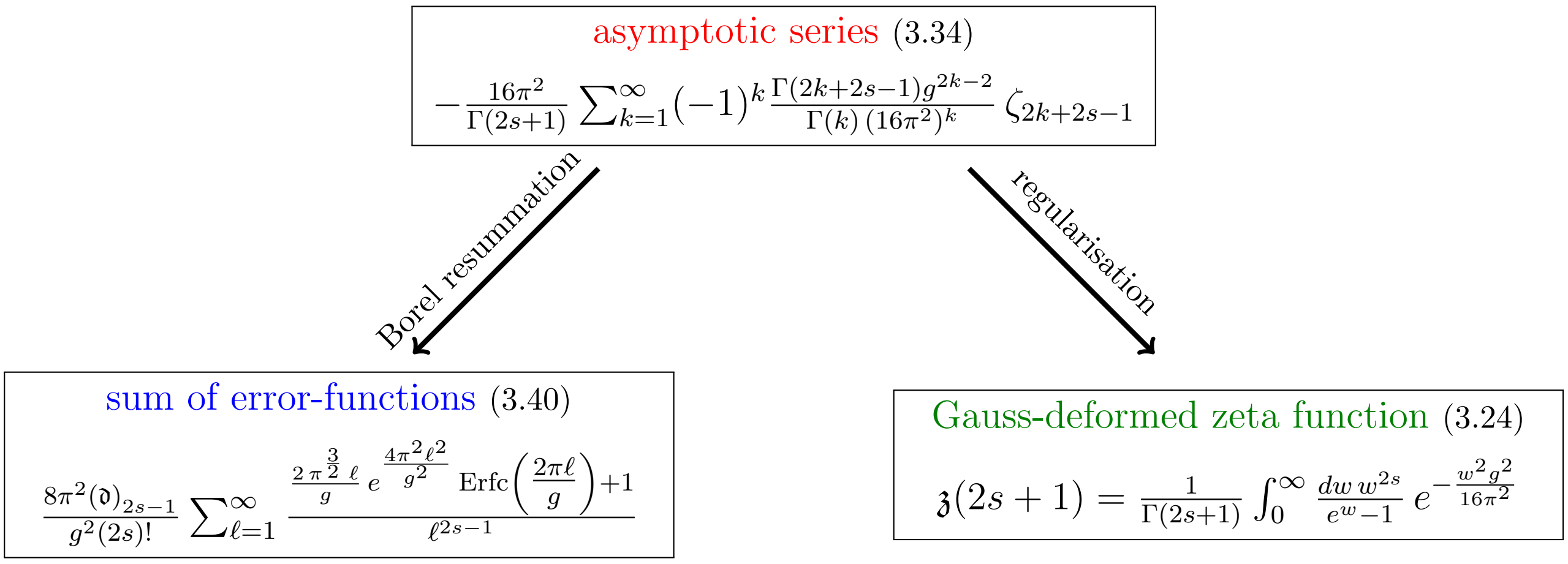

We stress that the two approaches are a priori equivalent but complementary. Indeed, in the concrete example of $\mathfrak{z}(2s+1)$ (for $s\in\mathbb{Z}_{>0}$) we work out an equivalence with a (converging) sum of error-functions as different ways of dealing with an asymptotic series, as is schematically shown in Figure~\ref{Fig:RelationAsymSeries}. 

However, writing the $\mathcal{Z}_\emptyset^{(2s),r}$ with the help of the $\mathfrak{z}$-functions leads generally to a much more compact presentation, which affords us a number of observations on the internal structure of the (perturbative part of the) partition function. Indeed, while the Borel resummation generally leads to infinite (albeit convergent) sums of error functions, the regularisation allows to express $\mathcal{Z}_\emptyset^{(2s),r}$ as combinations of finitely many of the $\mathfrak{z}$-functions (and standard Riemann zeta functions) in a way which respects the depth. Schematically, we find expressions of the form
\begin{align}
&\mathcal{Z}_\emptyset^{(2s),r}= \sum_{k=0}^r\mathcal{Z}_{\emptyset,k}^{(2s),r}\,,&&\text{with}&&\mathcal{Z}_{\emptyset,k}^{(2s),r}\sim \sum C(g) \underbrace{\zeta\ldots\zeta}_{(r-k)\text{ factors}} \times \underbrace{\mathfrak{z}(\ldots)}_{k\text{-arguments}}\,,\label{CartoonContrIntro}
\end{align}
with suitable coefficients $C(g)$ and the sum in $\mathcal{Z}_{\emptyset,k}^{(2s),r}$ runs only over finitely many terms (which we only indicate schematically). Our computations up to depth $3$ suggest recursive relations among the $\mathcal{Z}_{\emptyset,k}^{(2s),r}$ at different depths. In fact, we conjecture that at given $r$, only the term $\mathcal{Z}_{\emptyset,r}^{(2s),r}$, which contains only $\mathfrak{z}(n_1,\ldots,n_r)$ with the maximal number of arguments, provides new information, while all remaining terms can be fixed in terms of contributions at lower depth. 

Furthermore, at given depth, the $\mathfrak{z}(n_1,\ldots,n_r)$ for different arguments $(n_1,\ldots,n_r)$, are not independent but satisfy non-trivial relations among them (involving also derivatives with respect to $g$). These relations can be exploited to further simplify contributions to the partition function: we have demonstrated this possibility at depth $r=1$ and $r=2$ and found very compact expressions for $\mathcal{Z}_\emptyset^{(2)}$ and $\mathcal{Z}_\emptyset^{(4)}$. These simplifications remind of a similar situation in the study of multi-zeta value, where non-trivial relations are provided by shuffle- and stuffle relations. A more in-depth analysis of the relation to the latter is beyond the scope of the current work.

The remainder of this paper is organised as follows: In Section~\ref{Sect:IntRepPart} we provide more details on the partition function of $\mathcal{N}=2^*$ SYM theory on $\mathbb{S}^4$ and on computing the integral in (\ref{eq:intro_sphere_localization}) using the Laplace method. In Section~\ref{Sect:MassOrder}, we discuss the mass expansion of the perturbative contribution $\mathcal{Z}_\emptyset$ of the partition function: we use Borel resummation and a regularisation technique to give explicit expressions to general mass order $\mathcal{O}(\madj^{2s})$ at depth $r=1,2,3$. In Section~\ref{Sect:MassCorrection} we discuss in more detail the leading mass orders $\mathcal{Z}_\emptyset^{(2)}$ and $\mathcal{Z}_\emptyset^{(4)}$. Finally, Section~\ref{Sect:Conclusions} contains a summary of our Results and our Conclusions. This paper is accompanied by 2 Appendices, which give additional details and definitions of special functions used throughout the main text of the text and a brief review of Resurgence, respectively. 


\section{Integral Representation of the Partition Function}\label{Sect:IntRepPart}
We start by specifying a more detailed notation for the partition function (\ref{eq:intro_sphere_localization}) of $\mathcal{N}=2$ gauge theories on $\mathbb{S}^4$. For a more detailed discussion, we refer the reader to \cite{Pestun:2007rz,Hama:2012bg} (see also \cite{Pestun:2016zxk,Pestun:2016jze}). 
We set the radius of $\mathbb{S}^4$ unity in the following discussion.

\subsection{Four-dimensional Perturbative Partition Function on $\mathbb{S}^4$}

We consider supersymmetric gauge theories with the gauge group $G = U(n)$.
In this case, the Cartan subalgebra is identified with $\mathfrak{t} = i \mathbb{R}^n$.
Then, the classical part can be written in the form of a Gaussian factor and the integral can be written more explicitly 
\begin{align}
    \mathcal{Z}_{\mathbb{S}^4} = \int_{\mathbb{R}^n} [da] e^{-\rho \sum_i a_i^2} \,Z_{1\text{-loop}}\,\left| Z_{\rm inst}(ia,\madj)\right|^2\,,&&\text{with}&&\rho=\frac{8\pi^2}{g^2}\,,\label{PartitionIntegralRep}
\end{align}
where $Z_{\rm inst}$ denotes the Nekrasov instanton partition function \cite{Nekrasov:2002qd} with the self-dual $\Omega$-background, $\varepsilon_1 = - \varepsilon_2 = 1$. Furthermore, $\madj$ denotes the adjoint mass parameter, and $g$ is the dimensionless gauge coupling. The integral measure in (\ref{PartitionIntegralRep}) is given by 
\begin{align}
&[da]=\prod_{1\le\alpha<\beta\le n}\,(a_\alpha - a_\beta)^2\,\prod_{\alpha=1}^n\,d{a}_{\alpha}\,.
\end{align}
We remark that both the measure $[da]$ and the integrand in (\ref{PartitionIntegralRep}) only depend on the differences $a_\alpha -a_\beta$ corresponding to the roots associated with the gauge group $G$. Specifically, for $G = U(2)$ (in which case $\alpha,\beta\in\{1,2\}$), we can therefore rewrite the integral (\ref{PartitionIntegralRep}) in the new variable $a=a_2-a_1$
\begin{align}
 \mathcal{Z}_{\mathbb{S}^4} = \int_{\mathbb{R}} d\nu[a]\,\left| Z_{\rm inst}(ia,\madj)\right|^2\,,&&\text{with}&&d\nu[a]  = \frac{d\nu'[a]}{\int d\nu'[a]  }\,.\label{SpecificInstU2}
 \end{align}
Here we have defined the integral measure
\begin{align}
d\nu'[a]=da\,a^2\,e^{-\rho a^2/2}\,\Delta_{\text{1-loop}}[a,\madj]&&\text{for} &&a\in\mathbb{R}\,.\label{IntegrationMeasure}
\end{align}
with the 1-loop determinant
\begin{align}
&\Delta_{\text{1-loop}}[a,\madj]=\frac{e^{2(1+\gamma)\,\madj^2}\,H^2(\sqrt{2}\,i\,a)}{H(i(\sqrt{2}\,a+\madj))\,H(i(\sqrt{2}\,a-\madj)}\,,&&\text{with} &&H(z)=\mathsf{G}(1+z)\,\mathsf{G}(1-z)\,,\label{OneLoopDet}
\end{align}
where $\mathsf{G}$ is the Barnes $\mathsf{G}$-function~\cite{Barnes} (see (\ref{DefBarnesDoubleGamma}) in Appendix~\ref{App:Functions}) and $\gamma$ is the Euler-Mascheroni constant.\footnote{Relative to \cite{Pestun:2007rz}, we have  dropped a (divergent) $a$-independent factor in $\Delta_{\text{1-loop}}$, which is absorbed in the normalisation of the integral measure.} 
In this paper, for the sake of simplicity, we shall only be interested in the perturbative part of the partition function, \emph{i.e.} $Z_{\text{inst}}$ in (\ref{SpecificInstU2}) only contributes a constant, which we normalise to 1.

\subsection{Integral Measure and Laplace's Method}\label{Sect:LaplaceMethod}
In the following, we shall study the $U(2)$ perturbative partition function  
\begin{align}\label{eq:Z0}
 \mathcal{Z}_\emptyset(g,\madj)=\mathcal{Z}_\emptyset=\int_{\mathbb{R}}\,d\nu[a] \sim \int_{\mathbb{R}} da\,a^2\,e^{-\rho a^2}\,\Delta_{\text{1-loop}}[a,\madj] .
\end{align}
Notably due to the Gaussian factor $e^{-\rho\,a^2}$, the integrand is strongly peaked around $a=0$. We shall therefore evaluate $\mathcal{Z}_\emptyset$ using Laplace's method (see \emph{e.g.} \cite{BenderOrszag}). 

\subsubsection{Integral Measure and Series Expansion}\label{SeriesExpansionsMeasure}
To this end, we consider in more detail the 1-loop determinant in (\ref{OneLoopDet}). For small enough $a$, the Barnes $\mathsf{G}$-function allows the Taylor series expansion 
where $\gamma$ is the Euler-Mascheroni constant. We thus have
\begin{align}
H(z)&=\text{exp}\left[-(1+\gamma)z^2+\sum_{n=3}^\infty\frac{(-1)^{n-1}\zeta_{n-1}}{n}\,(z^n+(-z)^n)\right]\nonumber\\
&=e^{-(1+\gamma)z^2}\,\text{exp}\left(-\sum_{k=1}^\infty\frac{\zeta_{2k+1}}{k+1}\,z^{2k+2}\right)\,,&&\text{for} &&|z|<1\,.
\end{align}
and can therefore write 
\begin{align}
&\Delta_{\text{1-loop}}[a,\madj]
=\text{exp}\left(\sum_{k=1}^\infty\frac{2\zeta_{2k+1}}{k+1}\,\sum_{\ell=1}^{k+1}\left(\begin{linesmall}{@{}c@{}} \\[-6pt]2k+2 \\[3pt] 2\ell \end{linesmall}\right)\,(i\madj)^{2\ell}(\sqrt{2}ia)^{2k+2-2\ell}\right)\,,&&\text{for} &&-\tfrac{1}{\sqrt{2}}<a<\tfrac{1}{\sqrt{2}}\,.
\end{align}
We can furthermore neglect the term $\ell=k+1$, since it is an $a$ independent factor, which is absorbed in the normalisation of the partition function 
\begin{align}
&\Delta_{\text{1-loop}}[a,\madj]\sim\text{exp}\left(\sum_{k=1}^\infty\frac{2\zeta_{2k+1}}{k+1}\,\left((-1)^{k+1}(2k+2)(2k+1)\,a^2\,\madj^{2k}+\sum_{\ell=1}^{k-1}\left(\begin{linesmall}{@{}c@{}} \\[-6pt]2k+2 \\[3pt] 2\ell \end{linesmall}\right)\,(i\madj)^{2\ell}(\sqrt{2}ia)^{2k+2-2\ell}\right)\right)\nonumber\\
&=\text{exp}\left(c_2\, a^2+\sum_{k=1}^\infty\frac{2\zeta_{2k+1}}{k+1}\sum_{\ell=1}^{k-1}\left(\begin{linesmall}{@{}c@{}} \\[-6pt]2k+2 \\[3pt] 2\ell \end{linesmall}\right)\,(i\madj)^{2\ell}(\sqrt{2}ia)^{2k+2-2\ell}\right)\hspace{0.3cm}\text{for} \hspace{0.3cm}-\tfrac{1}{\sqrt{2}}<a<\tfrac{1}{\sqrt{2}}\,.\label{DeltaFormExpansion}
\end{align}
Here we have introduced the shorthand notation 
\begin{align}
c_2(\madj)&=4\sum_{k=1}^\infty(-1)^{k+1}(2k+1)\,\zeta_{2k+1}\,\madj^{2k}\nonumber\\
&=2\left(\madj\tfrac{d}{d\madj}+1\right)\left(2\gamma+\psi(1+i\madj)+\psi(1-i\madj)\right)\,,\hspace{0.5cm}\text{for} \hspace{0.5cm} |\madj|<1\,.\label{DefXiFun}
\end{align}
where $\psi$ is the digamma function (see (\ref{Digamma})), which, for $|z|<1$, allows the series expansion (\ref{SeriesExpansionDigamma}). Notice that $c_2(\madj)\geq 0$ for $-1<\madj<1$ and $c_2(\madj=0)=0$. For further use, we also rewrite
\begin{align}
\sum_{k=1}^\infty\frac{2\zeta_{2k+1}}{k+1}\sum_{\ell=1}^{k-1}\left(\begin{linesmall}{@{}c@{}} \\[-6pt]2k+2 \\[3pt] 2\ell \end{linesmall}\right)\,(i\madj)^{2\ell}(\sqrt{2}ia)^{2k+2-2\ell}=\sum_{s=2}^\infty c_{2s}(\madj)\, a^{2s}
\end{align}
with the coefficient functions
\begin{align}
&c_{2s}(\madj)=(-1)^s\,2^{s+1}\,\sum_{k=s}^\infty \frac{\zeta_{2k+1}}{k+1}\,\left(\begin{linesmall}{@{}c@{}} \\[-6pt]2k+2 \\[3pt] 2k+2-2s \end{linesmall}\right)\,(i\madj)^{2k+2-2s}\,,&&\forall s\geq 2\,,\label{FormckCoef}
\end{align}
for which we find explicitly 
\begin{align}
&c_{2s}(\madj)=(-1)^{s+1}\frac{2^{s+1}}{s}\,\zeta_{2s-1}\nonumber\\
&+2(-1)^{s+1}\frac{(2s-1)(\psi^{(2s-2)}(1+i\madj)+\psi^{(2s-2)}(1-i\madj))+i\madj\left(\psi^{(2s-1)}(1+i\madj)-\psi^{(2s-1)}(1-i\madj)\right)}{s!\,(2s-1)!!}\,.\nonumber
\end{align}
Close to the origin (\emph{i.e.} for $a\sim 0$), the integral measure (\ref{IntegrationMeasure}) then takes the form 
\begin{align}
&d\nu[a]=a^2\,e^{-a^2 \left(\rho-c_2(\madj)\right) }\,\text{exp}\left(\sum_{s=2}^\infty c_{2s}(\madj)\,a^{2s}\right)\,,&&\text{for} && -\tfrac{1}{\sqrt{2}}<a<\tfrac{1}{\sqrt{2}}\,.
\end{align}
We can further develop this measure by expanding the second exponential factor
\begin{align}
\text{exp}\left(\sum_{s=2}^\infty c_{2s}\,a^{2s}\right)=1+\sum_{k=2}^\infty d_{2k}(\madj)\,a^{2k}\,.
\end{align}
such that the integral measure takes the following form
\begin{align}
&d\nu[a]=a^2\,e^{-a^2 \left(\rho-c_2(\madj)\right) }\,\left(1+\sum_{k=2}^\infty d_{2k}(\madj)\,a^{2k}\right)\,,&&\text{for} && -\tfrac{1}{\sqrt{2}}<a<\tfrac{1}{\sqrt{2}}\,.\label{MeasureFactor}
\end{align}
which is a Gaussian factor multiplying a power series in $a$. In order to extract the coefficients $d_{2k}$ we consider
\begin{align}
&\sum_{n=1}^{\lceil k/2\rceil}\frac{1}{n!}\,\left(\sum_{s=2}^k c_{2s}\,a^{2s}\right)^n\nonumber\\
&\hspace{0.5cm}=\sum_{n=1}^{\lceil k/2\rceil}\frac{a^{4n}}{n!}\,\sum_{\ell_1=0}^n\sum_{\ell_2=0}^{\ell_1}\ldots \sum_{\ell_{k-2}=0}^{l_{k-3}}\left(\begin{linesmall}{@{}c@{}} \\[-6pt]n \\[3pt] \ell_1 \end{linesmall}\right)\left(\begin{linesmall}{@{}c@{}} \\[-6pt]\ell_1 \\[3pt] \ell_2 \end{linesmall}\right)\ldots \left(\begin{linesmall}{@{}c@{}} \\[-6pt]\ell_{k-3} \\[3pt] \ell_{k-2} \end{linesmall}\right)\,a^{2(\ell_1+\ldots+\ell_{k-2})}\,c_4^{n-\ell_1}\,c_6^{\ell_1-\ell_2}\ldots c_{2k}^{l_{k-2}}\,.
\end{align}
We thus have for the coefficient $d_{2k}$
\begin{align}
d_{2k}=\sum_{{\lceil k/2\rceil\geq n\geq \ell_1\geq \ldots\geq \ell_{k-2}\geq 0}\atop{2n+\ell_1+\ldots+\ell_{k-2}=k}}\frac{c_4^{n-\ell_1}\,c_6^{\ell_1-\ell_2}\ldots c_{2k}^{l_{k-2}}}{(n-\ell_1)!(\ell_1-\ell_2)!\ldots(\ell_{k-3}-\ell_{k-2})!\ell_{k-2}!}\,.\label{FormDcoeffDecomp}
\end{align}
Notice that due to (\ref{DefXiFun}) and (\ref{FormckCoef}), to leading order in $\madj$ we have
\begin{align}
d_{2k}=c_{2k}+\frac{1}{2}\sum_{\ell=2}^{k-2}c_{2\ell}\,c_{2k-2\ell}+\frac{1}{6}\,\sum_{\ell=2}^{k-4}\sum_{n=2}^{k-\ell-2}c_{2\ell}\,c_{2n}\,c_{2k-2\ell-2n}+\mathcal{O}(\madj^8)\,.
\end{align}

\subsubsection{Laplace Method}

We recall that the form (\ref{IntegrationMeasure}) of the 1-loop determinant in (\ref{DeltaFormExpansion}) is only valid in the interval $-\tfrac{1}{\sqrt{2}}<a<\tfrac{1}{\sqrt{2}}$. Therefore, in order to use this expansion, we first approximate $\mathcal{Z}_\emptyset$ by restricting to an integral region in which (\ref{DeltaFormExpansion}) is valid, \emph{i.e.}
\begin{align}
\mathcal{Z}_{\emptyset}\sim\int_{-\delta}^\delta da\,a^2\,e^{-\rho a^2}\,\Delta_{\text{1-loop}}[a,\madj]\,,\label{IntFormGenReduced}
\end{align}
for some $0<\delta<\frac{1}{\sqrt{2}}$. Following \cite{BenderOrszag}, this is correct up to terms that are exponentially suppressed for $\rho\to\infty$, \emph{i.e.} instanton contributions in the weak coupling limit. For the new integral in (\ref{IntFormGenReduced}), we can use the expansions (\ref{MeasureFactor}) explained in Section~\ref{SeriesExpansionsMeasure}). Individual terms in the expansion now give rise to Gaussian integrals of the form
\begin{align}
\int_{-\delta}^\delta a^{2+2k}\,e^{-(\rho-c_2(\madj))a^2}\,d a\sim \int_{-\infty}^\infty a^{2+2k}\,e^{-(\rho-c_2(\madj))a^2}\,d a= \frac{\Gamma(k+3/2)}{(\rho-c_2(\madj))^{k+3/2}}\,.\label{GaussIntegr}
\end{align}
This is once again correct up to terms that are exponentially suppressed for $\rho\to\infty$, \emph{i.e.} instanton corrections.

\section{Generic Mass Expansion of the Perturbative Contribution}\label{Sect:MassOrder}
Following the discussion of the previous Section, we are lead to study the following expression for the perturbative contribution of the partition function
\begin{align}
\mathcal{Z}_{\emptyset}=\int_{\mathbb{R}}da\,a^2\,e^{-a^2 \left(\rho-c_2(\madj)\right) }\,\left(1+\sum_{k=2}^\infty d_{2k}(\madj)\,a^{2k}\right)\,.
\end{align}
Using the approximation (\ref{GaussIntegr}), we can write $\mathcal{Z}_{\emptyset}$ as the following series
\begin{align}
\mathcal{Z}_{\emptyset}=\frac{\sqrt{\pi}}{2(\rho-c_2(\madj))^{3/2}}+\sum_{k=2}^\infty d_{2k}\, \frac{\Gamma(k+3/2)}{(\rho-c_2(\madj))^{k+3/2}}\,.\label{Z00sumD}
\end{align}
Using furthermore the decomposition (\ref{FormDcoeffDecomp}), we can write 
\begin{align}
\mathcal{Z}_{\emptyset}=\sum_{r=0}^\infty \mathcal{Z}_{\emptyset}^r\,,\label{DecompDepth}
\end{align}
where $\mathcal{Z}_{\emptyset}^r$ contains exactly $r$ factors $c_{2a}$ (for $a\geq 1$). Concretely, we have for the first three terms
{\allowdisplaybreaks
\begin{align}
\mathcal{Z}_{\emptyset}^0&=\frac{\sqrt{\pi}}{2\rho^{3/2}}\,,\nonumber\\
\mathcal{Z}_{\emptyset}^1&=\frac{3\sqrt{\pi}}{4\rho^{5/2}}\,c_2+\sum_{k=2}^\infty c_{2k}\,\frac{\Gamma(k+3/2)}{\rho^{k+3/2}}=\sum_{k=1}^\infty c_{2k}\,\frac{\Gamma(k+3/2)}{\rho^{k+3/2}}\,,\nonumber\\
\mathcal{Z}_{\emptyset}^2&=\frac{15\sqrt{\pi}}{16\,\rho^{7/2}}\,c_2^2+\frac{1}{2}\sum_{k=2}^\infty\left(\frac{(2k+3)}{\rho}c_{2k}\,c_2+\sum_{\ell=2}^{k-2}c_{2\ell}\,c_{2k-2\ell}\right)\frac{\Gamma(k+3/2)}{\rho^{k+3/2}}\,,\nonumber\\
\mathcal{Z}_{\emptyset}^3&=\frac{35\sqrt{\pi}}{32\rho^{9/2}}\,c_2^3+\sum_{k=2}^\infty\bigg[\frac{(2k+3)(2k+5)}{8\rho^{2}}\,c_2^2\,c_{2k}+\frac{2k+3}{4\rho}\,c_2\,\sum_{\ell=2}^{k-2}c_{2\ell}\,c_{2k-2\ell}\nonumber\\
&\hspace{3.75cm}+\frac{1}{6}\,\sum_{\ell=2}^{k-4}\sum_{n=2}^{k-\ell-2}c_{2\ell}\,c_{2n}\,c_{2k-2\ell-2n}\bigg]\,\frac{\Gamma(k+3/2)}{\rho^{k+3/2}}\,.\label{ZdepthDecomposition}
\end{align}}
An individual term of the form $\mathcal{Z}_{\emptyset}^r$ only provides contributions to order $\mathcal{O}(\madj^{2s})$ with $s\geq r$
\begin{align}
\mathcal{Z}_{\emptyset}^r=\sum_{s=r}^\infty \madj^{2s}\,\mathcal{Z}_{\emptyset}^{(2s),r}
\end{align}
such that the coefficients $\mathcal{Z}_{\emptyset}^{(2s)}$ in the mass-expansion of $\mathcal{Z}_\emptyset$ take the following decomposition
\begin{align}
&\mathcal{Z}_{\emptyset}=\sum_{s=0}^\infty \madj^{2s}\,\mathcal{Z}_{\emptyset}^{(2s)}\,,&&\text{with} &&\mathcal{Z}_{\emptyset}^{(2s)}=\sum_{r=1}^s\mathcal{Z}_{\emptyset}^{(2s),r}\,.\label{IntroZ2sr}
\end{align}
In the following we shall study the leading $\mathcal{Z}_{\emptyset}^{(2s),r}$ for $r=1,2,3$. From the definition (\ref{ZdepthDecomposition}), these are infinite sums over $k$, which, however, generally have vanishing radius of convergence. In order to make sense of these expressions, we shall explore two different approaches: 
\begin{itemize}
\item[\emph{(i)}] {\bf Borel resummation:} the series are in fact asymptotic series, which can be resummed into infinite (convergent) series of error-functions. A similar approach has been employed at various other cases in the literature (see notably the recent work \cite{Aniceto:2026pcz})
\item[\emph{(ii)}] {\bf regularisation:} based on the presence of the Riemann zeta function in the coefficients $c_{2n}$ in (\ref{FormckCoef}), we shall propose a regularisation, which renders the summation over $k$ convergent: the result can be cased in the form of \emph{finite} combinations of Gauss-deformed zeta functions, which we shall introduce.
\end{itemize}
The presence of Riemann zeta functions is crucial for the approach \emph{(ii)}. Since the general term $\mathcal{Z}_{\emptyset}^{(2s),r}$ is defined as the contribution to the partition function constructed from combinations of precisely $r$ coefficients $c_{2n}$ in (\ref{FormckCoef}), it also corresponds to combinations of $r$ Riemann zeta functions. Borrowing the convention of the literature of multi-zeta values (see \emph{e.g.} \cite{Brown:2011ik}) we shall therefore refer to $\mathcal{Z}_{\emptyset}^{(2s),r}$ as the contribution of \emph{depth} $r$ to the term of order $\mathcal{O}(\madj^{2s})$ to the partition function. 
\subsection{Contribution of Depth 1}\label{Sect:Depth1Contribution}
We start with the contribution $\mathcal{Z}_\emptyset^1$ in (\ref{ZdepthDecomposition}), which captures all contributions that contain only one factor of $c_{2k}$ in (\ref{Z00sumD}). Using the form of $c_{2k}$ in (\ref{DefXiFun}) and (\ref{FormckCoef}), we find for the contribution of order $\mathcal{O}(\madj^{2s})$
\begin{align}
\mathcal{Z}_{\emptyset}^{(2s),1}&=(-1)^s\sum_{k=1}^\infty(-1)^k\frac{\Gamma(k+3/2)}{\rho^{k+3/2}}\,2^{k+1}\frac{\zeta_{2k+2s-1}}{k+s}\,\big(^{2k+2s}_{\hspace{0.3cm}2s}\big)\,.\label{Depth1Series}
\end{align}
This series (as a function of $g$) has zero radius of convergence. Indeed, using the identity
\begin{align}
 \Gamma(k+3/2)=\frac{(2k+2)!\,\sqrt{\pi}}{4^{k+1}(k+1)!}=\frac{2(2k+1)\sqrt{\pi}}{4^{k+1}}\frac{(2k)!}{k!}\,,
\end{align}
we may equivalently write (\ref{Depth1Series}) in the following form
\begin{align}
\mathcal{Z}_{\emptyset}^{(2s),1}&=\frac{2\sqrt{\pi}(-1)^s}{\rho^{3/2}}\sum_{k=1}^\infty\frac{(-1)^k}{\rho^{k}}\,\frac{\zeta_{2k+2s-1}}{k+s}\,\frac{(2k+1)}{2^{k+1}}\frac{(2k)!}{k!}\,\big(^{2k+2s}_{\hspace{0.3cm}2s}\big)\nonumber\\
&=\frac{2\sqrt{\pi}(-1)^s}{(2s)!\rho^{3/2}}\sum_{k=1}^\infty\frac{(-1)^k\zeta_{2k+2s-1}}{(8\pi^2)^k\, 2^k}\,\frac{(2k)!}{k!}\,(2k+1)^2(2k+2s-1)\ldots(2k+2)\,g^{2k}\,.\label{InitialBorel}
\end{align}
In order to still make sense of this expression, we may use the presence of the Riemann zeta function $\zeta_{2k+2s-1}$ in two different fashions, namely to Borel resum (\ref{InitialBorel}) or to regularise the summation over $k$. 

\subsubsection{Borel Resummation}\label{Sect:BorelResumFirst}
As a first approach, we use the definition (\ref{DefZetaFunction}) of the Riemann zeta function to write (\ref{InitialBorel}) as
\begin{align}
\mathcal{Z}_{\emptyset}^{(2s),1}=\frac{2\sqrt{\pi}(-1)^s}{(2s)!\rho^{3/2}}\sum_{\ell=1}^\infty\frac{1}{\ell^{2s-1}}\sum_{k=1}^\infty\frac{(-1)^k}{ (16\pi^2)^k\,\,\ell^{2k}}\,\frac{(2k)!}{k!}\,(2k+1)\,g^{2k}\prod_{a=1}^{2s-1}(2k+a)\,.
\end{align}
We furthermore express the factors $(2k+a)$ by using the identity 
\begin{align}
\left(g\,\tfrac{d}{dg}+a \right)\,g^{2k}=(\dop+a)\,g^{2k}=(2k+a)\,g^{2k}\,.\label{DiffOperatorID}
\end{align}
In this way, we formally obtain the following rewriting of (\ref{InitialBorel})
\begin{align}
\mathcal{Z}_{\emptyset}^{(2s),1}=\frac{2\sqrt{\pi}(-1)^s}{(2s)!\rho^{3/2}}\sum_{\ell=1}^\infty\frac{1}{\ell^{2s-1}}\left(\dop+1\right)\,\left(\dop+1\right)_{2s-1}\,\ser\left(\tfrac{g^2}{\ell^2}\right)\,.\label{SingleDepthGenBuildingBlock}
\end{align}
Here we have used as a shorthand notation the Pochhammer symbol defined in (\ref{DefPochhammer}), \emph{i.e.} $\left(\dop+1\right)_{2s-1}=\prod_{a=1}^{2s-1}\left(g\,\tfrac{d}{dg}+a\right)$. The expression (\ref{SingleDepthGenBuildingBlock}) is still divergent, since the series 
\begin{align}
&\ser(z):= \sum_{k=1}^\infty \alpha_{k}\,z^{k}\,,&&\text{with} &&\alpha_{k}=(-1)^k\,\frac{(2k)!}{k!\, (16\pi^2)^k}\,,
\end{align}
is an asymptotic series with vanishing radius of convergence. More concisely, $\ser$ is of Gevrey-1 type \cite{Dorigoni:2014hea} and is studied in more detail in Appendix~\ref{App:BorelResumSer}. There, it is notably recalled (see also \cite{Aniceto:2026pcz}) that $\ser$ can be Borel resummed to give (\ref{BorelResumS1}), such that
\begin{align}
\mathcal{L}^0[\mathcal{B}[\ser]](g^2/\ell^2)=\frac{2\,\pi^{3/2}\,\ell}{g}\,e^{\frac{4\pi^2\ell^2}{g^2}}\,\left(1-\text{Erf}\left(\tfrac{2\pi\ell}{g}\right)\right)-1\,,\label{DefLaplaceBell}
\end{align}
where Erf is the error function (see~(\ref{DefErfFunction})). Indeed, the error-function affords an asymptotic expansion of the form (\ref{ErfExpansion}) (see also (\ref{SerExpErf})), which, when substituted into (\ref{DefLaplaceBell}) reproduces $\ser$ in (\ref{SingleDepthGenBuildingBlock}.) The appearance of the error function in this context has been observed previously in \cite{Aniceto:2026pcz,dorigoni2024electromagneticdualitylinedefect} (see also \cite{Witten:1992xu} for the appearance of error functions in the weak coupling expansion of 2-dimensional theories). We can therefore write
\begin{align}
\mathcal{Z}_{\emptyset}^{(2s),1}=\frac{2\sqrt{\pi}(-1)^s}{(2s)!\rho^{3/2}}\sum_{\ell=1}^\infty\frac{1}{\ell^{2s-1}}\left(\dop+1\right)\,\left(\dop+1\right)_{2s-1}\,\mathcal{L}^0[\mathcal{B}[\ser]](g^2/\ell^2)\,.\label{BorelResumSingleDepth}
\end{align}
While still an infinite series (over $\ell$), the latter is in fact convergent, as is for example concretely discussed for $s=1$ in Section~\ref{Sect:LeadingMassDeformation}.


\subsubsection{Regularisation}\label{Sect:Depth1Reg}
We return to (\ref{Depth1Series}) to calculate it in a different fashion. To this end, we first use the Mellin representation (\ref{FormZetaIntegSinh}) of the Riemann zeta functions\footnote{For $s=1$, this rewriting was previously advocated in \cite{Dorigoni:2021guq} (see also \cite{Chester:2019pvm}).}
\begin{align}
\mathcal{Z}_{\emptyset}^{(2s),1}&=(-1)^s\sum_{k=1}^\infty(-1)^k\frac{\Gamma(k+3/2)}{\rho^{k+3/2}}\,\frac{2^{k+1}}{k+s}\,\big(^{2k+2s}_{\hspace{0.3cm}2s}\big)\,\frac{2^{2k+2s-2}}{\Gamma(2k+2s)}\int_0^\infty\frac{dw\,w^{2k+2s-1}}{\sinh^2(w)}\nonumber\\
&=\frac{2^{2s-1}\sqrt{\pi}(-1)^s}{(2s)!\rho^{3/2}}\sum_{k=1}^\infty\frac{(-1)^k\,2^k}{\rho^k}\,\frac{(2k)!}{k!}\,\frac{2k+1}{\Gamma(2k+2s)}\left(\prod_{a=1}^{2s-1}(2k+a)\right)\int_0^\infty\frac{dw\,w^{2k+2s-1}}{\sinh^2(w)}\,.
\end{align}
The series with general term 
\begin{align}
\frac{(-1)^k\,2^k}{\rho^k}\,\frac{(2k)!}{k!}\,\frac{2k+1}{\Gamma(2k+2s)}\left(\prod_{a=1}^{2s-1}(2k+a)\right)\,w^{2k+2s-1}
\end{align}
converges uniformly on any closed interval of the form $[0,x]$ for any finite $0<x<\infty$, but not on the interval $[0,\infty)$. Thus, in order to be able to exchange the summation with the integration, we \emph{regularise} $\mathcal{Z}_{\emptyset}^{(2s),1}$ by defining
\begin{align}
\mathcal{Z}_{\emptyset,\delta}^{(2s),1}&=\frac{2^{2s-1}\sqrt{\pi}(-1)^s}{(2s)!\rho^{3/2}}\sum_{k=1}^\infty\frac{(-1)^k\,2^k}{\rho^k}\,\frac{(2k)!}{k!}\,\frac{2k+1}{\Gamma(2k+2s)}\left(\prod_{a=1}^{2s-1}(2k+a)\right)\int_0^{\frac{1}{2\delta}}\frac{dw\,w^{2k+2s-1}}{\sinh^2(w)}\nonumber\\
&=\frac{2^{2s-1}\sqrt{\pi}(-1)^s}{(2s)!\rho^{3/2}}\int_0^{\frac{1}{2\delta}}\frac{dw\,w^{2s-1}}{\sinh^2(w)}\sum_{k=1}^\infty(-1)^k\,\frac{(2k)!}{k!}\,\left(\frac{2w^2}{\rho}\right)^{k}\,\frac{2k+1}{\Gamma(2k+1)}\,,\label{Z2s1regularisation}
\end{align}
for a suitable regularisation parameter $\delta\ll 1$. The remaining summation over $k$ is independent of $s$ and leads to
\begin{align}
\mathcal{Z}_{\emptyset,\delta}^{(2s),1}&=\frac{2^{2s-1}\sqrt{\pi}(-1)^s}{(2s)!\rho^{5/2}}\int_0^{\frac{1}{2\delta}}\frac{dw\,w^{2s-1}}{\sinh^2(w)}\left(e^{-\frac{2w^2}{\rho}}(\rho-4w^2)-\rho\right)\,.
\end{align}
For further convenience, we shall rescale $w\to w/2$, leading to
\begin{align}
\mathcal{Z}_{\emptyset,\delta}^{(2s),1}&=\frac{\sqrt{\pi}(-1)^s}{2(2s)!\rho^{5/2}}\int_0^{1/{\delta}}\frac{dw\,w^{2s-1}}{\sinh^2(w/2)}\left(e^{-\frac{w^2}{2\rho}}(\rho-w^2)-\rho\right)\,.
\end{align}
Using the identity
\begin{align}
&\frac{1}{\sinh^2(x/2)}=-\frac{d}{dx}\,\frac{4}{e^x-1}\,,&&\forall x\in\mathbb{R}_+\,,\label{ExpForm}
\end{align}
we have after an integration by parts
\begin{align}
\mathcal{Z}_{\emptyset,\delta}^{(2s),1}&\sim\frac{2\sqrt{\pi}(-1)^s}{(2s)!\rho^{5/2}}\int_0^{1/{\delta}}d^1[w]\,w^{2s-2}\,\left(e^{-\frac{w^2}{2\rho}}\left(\frac{w^4}{\rho}-2(s+1)\,w^2+(2s-1)\rho\right)-(2s-1)\rho\right)\,,\label{ResZeta2s1}
\end{align}
where $\int_0^{1/\delta}d^1[w]:=\int_0^{1/\delta}\frac{dw}{e^w-1}$ and $\sim$ denotes equality up to terms which vanish in the limit $\delta\to 0$.
This is in fact a convergent integral and we therefore write for the regularised contribution of order $\mathcal{O}(\madj^2)$ to the partition function
\begin{align}
\mathcal{Z}_{\emptyset}^{(2s),1}=\frac{(-1)^s\sqrt{\pi}}{\rho^{7/2}}\left[\frac{\rho^2}{s}(\ztu{2s-1}-\zeta_{2s-1})-4(s+1)\,\rho\,\ztu{2s+1}+4(s+1)(2s+1)\,\ztu{2s+3}\right]\,.\label{SingleDepthZetaFunction}
\end{align}
Here we have introduced \emph{Gauss-deformed zeta functions} as the following Mellin integrals with a Gaussian kernel
\begin{align}
\ztu{a}=\frac{1}{\Gamma(a)}\int_0^\infty\frac{e^{-\frac{t^2}{2\rho}}\,t^{a-1}}{e^t-1}\,dt\,,&&\forall a>1\,.\label{SingleZetReg}
\end{align}
Although Mellin representations of zeta functions with similar kernels have been discussed in the mathematics literature (see notably~\cite{hansen2026parabolicmellintransformgamma,ALBEVERIO200712}), to the best of our knowledge, the appearance of the functions (\ref{SingleZetReg}) in the current physics context has not been previously proposed in the literature. Notice, if (inspired by the literature on multi-zeta values) we define the argument of a $\mathfrak{z}$-function as its weight, the latter is correlated with the explicit power of $\rho$ appearing in the square bracket in (\ref{SingleDepthZetaFunction}): indeed, the terms with power $\rho^\ell$ have weight $2s-3-2\ell$ (for $\ell=0,1,2$). We also note in passing that $\ztu{a}$ satisfies the following recursive relation 
\begin{align}
&\ztu{a+2}=-\frac{\rho}{a(a+1)}\,g\,\frac{d}{dg}\,\ztu{a}=:-\frac{\rho}{a(a+1)}\,\dop\,\ztu{a}\,,&&\text{with} &&\dop:=g\,\frac{d}{dg}\,,\label{ShiftZetaValue}
\end{align}
which allows to write (\ref{SingleZetReg}) in the more compact form
\begin{align}
\mathcal{Z}_{\emptyset}^{(2s),1}=\frac{(-1)^s\sqrt{\pi}}{\rho^{3/2}}\left[\frac{(\dop+1)(\dop+2s-1)}{(2s-1)s}\ztu{2s-1}-\frac{\zeta_{2s-1}}{s}\right]\,.\label{ZpartZeta}
\end{align}
Using this relation to express $\mathcal{Z}_{\emptyset}^{(2s+2),1}$ in the form of a differential operator acting on $\ztu{2s+1}$ and using the shift identity (\ref{ShiftZetaValue})
\begin{align}
\mathcal{Z}_{\emptyset}^{(2s+2),1}&=\frac{(-1)^s\sqrt{\pi}}{\rho^{3/2}}\left[\frac{\rho\,\dop(\dop-1)(\dop+2s-1)}{(2s-1)2s(2s+1)(s+1)}\,\ztu{2s-1}+\frac{\zeta_{2s+1}}{s+1}\right]
\end{align}
leads together with (\ref{ZpartZeta}) to the following recursive relation
\begin{align}
&\dop\,\mathcal{Z}_{\emptyset}^{(2s+2),1}=\frac{\rho\,(\dop-4)(\dop-3)}{(2s+2)(2s+1)}\,\mathcal{Z}_{\emptyset}^{(2s),1}+\frac{3(-1)^s\sqrt{\pi}}{\rho^{3/2}(s+1)}\,\zeta_{2s+1}\,,&&\forall s\geq 1\,.
\end{align}

The case $s=1$ in (\ref{SingleDepthZetaFunction}) demands a further remark: indeed for $s=1$, neither the Riemann zeta function $\zeta_{2s-1}$ nor $\ztu{2s-1}$ are well defined (since for $s=1$, the integral in (\ref{SingleZetReg}) diverges at $t\to 0$). However, the contribution of order $\mathcal{O}(\madj^2)$ to the partition function
\begin{align}
\mathcal{Z}_{\emptyset}^{(2),1}&=-\frac{\sqrt{\pi}}{\rho^{7/2}}\int_0^\infty d^1[w]\,\left(e^{-\frac{w^2}{2\rho}}\left(w^4-4w^2 \rho+\rho^2\right)-\rho^2\right)\,,\label{Z1FormInteg}
\end{align}
is still well-defined as integral at $w=0$ (since the first order pole cancels among the different contributions). In (\ref{SingleZetReg}), for $s=1$, which takes the form
\begin{align}
\mathcal{Z}_{\emptyset}^{(2)}&=-\frac{\sqrt{\pi}}{\rho^{7/2}}\left(\rho^2\,(\ztu{1}-\zeta_1)-8\,\rho\,\ztu{3}+24\,\ztu{5}\right)\,,\label{ZetaResultZ2}
\end{align}
we therefore understand the somewhat abusive notation
\begin{align}
\ztu{1}-\zeta_1=\int_0^\infty\,d^1[t]\,\left(e^{-\frac{t^2}{2\rho}}-1\right)\,,\label{AbusiveDefZeta1}
\end{align}
which is indeed well-defined. An alternative way of understanding the regularity of (\ref{ZetaResultZ2}) and to give more rigour to the notation (\ref{AbusiveDefZeta1}) is to regularise the integrals at the origin with a cutoff $\epsilon\ll 1$, \emph{i.e.}
\begin{align}
\zeta_1&\sim_{\epsilon\to 0}\int_\epsilon^\infty d^1[t]=-\log(\epsilon)+\mathcal{O}(\epsilon)\,,\nonumber\\
\mathfrak{z}(1)&\sim_{\epsilon\to 0} \int_\epsilon^\infty d^1[t]\,e^{-\frac{t^2}{2\rho}}=-\log(\epsilon)+\int_0^\infty\,d^1[t]\,\left(e^{-\frac{t^2}{2\rho}}-1\right)+\mathcal{O}(\epsilon)\label{RegularisationEpsilon}
\end{align} 
such that we have for their combination
\begin{align}
\mathfrak{z}(1)-\zeta_1\sim_{\epsilon\to 0}\int_0^\infty\,d^1[t]\,\left(e^{-\frac{t^2}{2\rho}}-1\right)\,.\label{EpsilonRegularisation}
\end{align}
This indeed reproduces (\ref{AbusiveDefZeta1}). In the following, we implicitly understand potentially divergent integrals to be regularised by appropriate cutoff parameters as above. 

\begin{figure}
\vspace{-0.25cm}
\begin{center}
\includegraphics[width = 7.5cm]{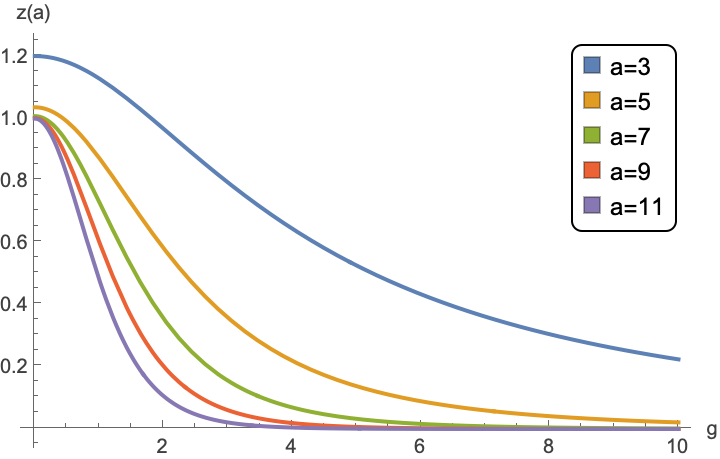}\hspace{1cm}\includegraphics[width = 7.5cm]{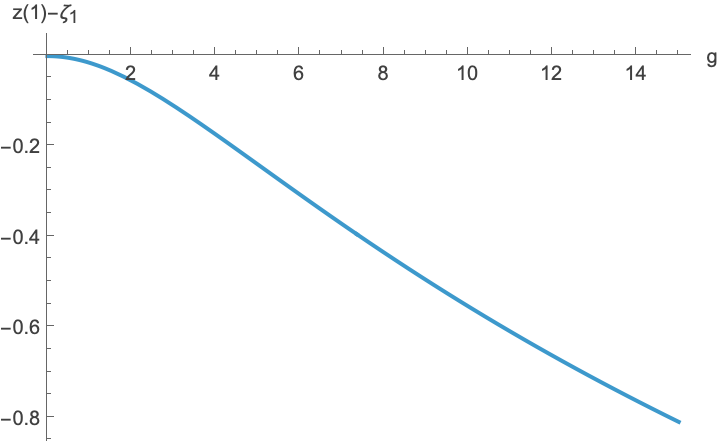}
\end{center}
\vspace{-0.5cm}
\caption{Left panel: Numerical plot of $\ztu{a}$ for odd values of $a\geq 3$ as a function of $g$. Right panel: $\ztu{1}-\zeta_1$ as symbolically defined in eq.~(\ref{AbusiveDefZeta1}.}
\label{Fig:NumericalSingleZeta}
\vspace{-0.25cm}
\end{figure}

Numerical visualisations of $\ztu{a}$ as well as (\ref{AbusiveDefZeta1}) are shown in Figure~\ref{Fig:NumericalSingleZeta}. A comparison between the Borel resummed expression (\ref{BorelResumSingleDepth}), the regularised result (\ref{SingleDepthZetaFunction}) and (truncations of) the initial asymptotic series (\ref{Depth1Series}) for $s=1$ are shown in Figure~\ref{Fig:NumericalZ1} in Section~\ref{Sect:MassCorrection}.

\subsubsection{Correspondence: $\mathfrak{z}$-Function as Asymptotic Series}\label{Sect:Correspondence}
Before moving on to contributions to the partition function of higher depth, we first comment on a non-trivial correspondence that is suggested by the above results. Indeed, comparing (\ref{SingleDepthZetaFunction}) with (\ref{BorelResumSingleDepth}) (using notably (\ref{DefLaplaceBell})) as well as (\ref{Depth1Series}), suggests a relation between the Gauss-deformed zeta functions we have introduced in (\ref{SingleZetReg}) with (inifnite series of) error-functions, both of which being represented by asymptotic series. In order to make this connection more concrete, we shall provide an asymptotic series, whose regularisation (along the lines of Section~\ref{Sect:Depth1Reg}) leads to $\ztu{2s+1}$ and whose Borel resummation can be written as an infinite sum of error-functions. Our starting point is 
\begin{align}
\mathcal{I}_s=-\frac{2\rho}{\Gamma(2s+1)}\sum_{k=1}^\infty (-1)^k\frac{\Gamma(2k+2s-1)}{\Gamma(k)\,2^k\,\rho^k}\,\zeta_{2k+2s-1}\,,\label{SeriesZetaSingle}
\end{align}
which is an asymptotic series, which we regularise by writing $\zeta_{2k+2s-1}$ in its Mellin representation (\ref{FormZetaIntegSinh}), for which we introduce a cutoff as in (\ref{Z2s1regularisation})
\begin{align}
\mathcal{I}_{s,\delta}=-\frac{2^{2s}}{(2s)!}\sum_{k=1}^\infty\frac{(-1)^k 2^{k-1}}{\rho^{k-1}(2k+2s-1)\Gamma(k)}\,\int_0^{\frac{1}{2\delta}}\,\frac{w^{2k+2s-1}\,dw}{\sinh^2(w)}\,.
\end{align}
Exchanging integration with summation, we obtain
\begin{align}
\mathcal{I}_{s,\delta}=-\frac{2^{2s}}{(2s)!}\int_0^{\frac{1}{2\delta}}\frac{w^{2s}\,dw}{\sinh^2(w)}\sum_{k=1}^\infty\frac{(-1)^k 2^{k-1} w^{2k-1}}{\rho^{k-1}(2k+2s-1)\Gamma(k)}\,.
\end{align}
Using (\ref{SeriesErrorFct}) we can explicitly perform the summation over $k$ to obtain (where we have also rescaled the integration variable $w\to w/2$) 
\begin{align}
\mathcal{I}_{s,\delta}=-\frac{1}{2(2s)!}\int_0^{\frac{1}{\delta}}\frac{w^{2s}\,dw}{\sinh^2(w/2)}\left[e^{-\frac{w^2}{2\rho}}\,\rho\,w\,\frac{(2s-1)!!}{2(2s)!}\sum_{\ell=0}^{s-1}\left(\frac{4\,\ell\, w^{2\ell}\,\rho^{s-\ell-1}}{2^{2\ell}(2\ell+1)!!}\right)-\frac{\sqrt{2\pi}\,\rho^{s+1/2}}{2^{s+2}\,s!}\,\text{Erf}\left(\tfrac{w}{\sqrt{2\rho}}\right)\right]\,.
\end{align}
After an integration by parts (using (\ref{ExpForm})) we can write 
\begin{align}
\mathcal{I}_{s,\delta}\sim\frac{1}{\Gamma(2s+1)}\int_0^{\frac{1}{\delta}}d^1[w]\,w^{2s}\,e^{-\frac{w^2}{2\rho}}\sim \ztu{2s+1}\,,
\end{align}
where $\sim$ indicates equality up to terms that vanish in the limit $\delta\to 0$. In the limit we then indeed find $\lim_{\delta\to 0} \mathcal{I}_{s,\delta}=\ztu{2s+1}$. The series in (\ref{SeriesZetaSingle}) therefore represents an asymptotic series, whose regularisation leads to the Gauss-deformed zeta functions defined in (\ref{SingleZetReg}).

\begin{figure}
\vspace{-0.25cm}
\begin{center}
\includegraphics[width = 7.5cm]{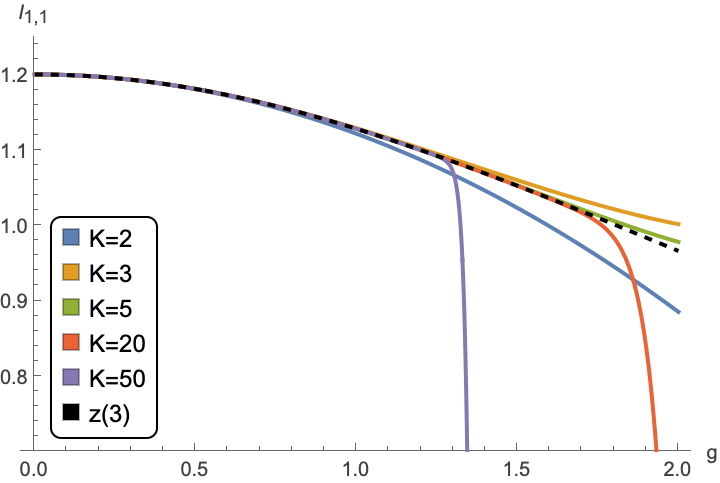}\hspace{1cm}\includegraphics[width = 7.5cm]{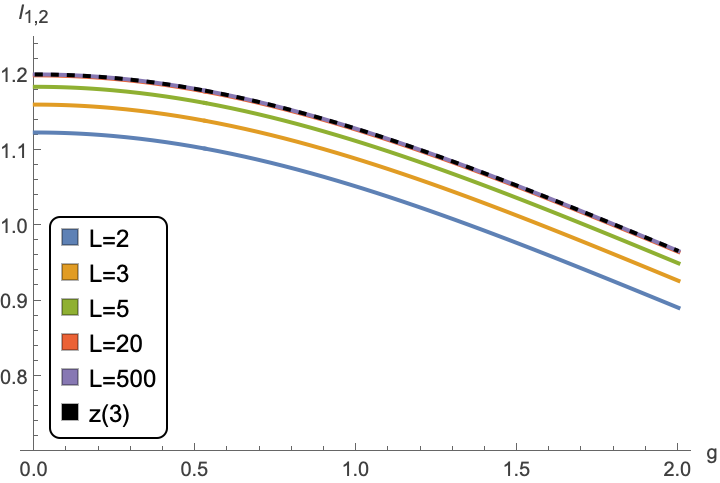}
\end{center}
\vspace{-0.5cm}
\caption{Graphical visualisation of $\mathcal{I}_1$. Left panel: comparison of (\ref{SeriesZetaSingleL}) for different values of the cutoff $K$; right panel: comparison of (\ref{SeriesBorelResumCorrL}) for different values of $L$. In both plots, the dashed black line represents (a numerical evaluation of) the Gauss-deformed zeta function $\ztu{3}$ in (\ref{SingleZetReg}). Notice in particular that at $g=0$ we have $\ztu{3}(g=0)=\zeta_3\sim1.20206$.}
\label{Fig:ResummationProcedureI1}
\vspace{-0.25cm}
\end{figure}

Alternatively to regularising the sum in (\ref{SeriesZetaSingle}), we can also Borel resum it in a similar fashion as in Section~\ref{Sect:BorelResumFirst}. To this end, we rewrite
\begin{align}
\mathcal{I}_s=-\frac{8\pi^2}{g^2(2s)!}\sum_{k=1}^\infty\left(2k\right)_{2s-1}\,\zeta_{2k+2s-1}\,\frac{(-1)^k (2k)!}{k! (16\pi^2)^k}\,g^{2k}\,.
\end{align}
with $(2k)_{2s-1}=\prod_{a=0}^{2s-2}(2k+a)$ (see (\ref{DefPochhammer})). Using (\ref{DefZetaFunction}) and (\ref{DiffOperatorID}), we may write
\begin{align}
\mathcal{I}_s&=-\frac{8\pi^2}{g^2(2s)!}\sum_{\ell=1}^\infty\frac{1}{\ell^{2s-1}}\sum_{k=1}^\infty\left(\dop\right)_{2s-1}\,\frac{(-1)^k (2k)!}{k! (16\pi^2)^k}\,\left(\frac{g^2}{\ell^2}\right)^{k}=-\frac{8\pi^2}{g^2(2s)!}\left(\dop\right)_{2s-1}\sum_{\ell=1}^\infty\frac{1}{\ell^{2s-1}}\,\ser\left(\tfrac{g^2}{\ell^2}\right)\,,\nonumber
\end{align}
which lends itself to the Borel re-summation
\begin{align}
\mathcal{I}_s&=-\frac{8\pi^2}{g^2(2s)!}\left(\dop\right)_{2s-1}\sum_{\ell=1}^\infty\frac{1}{\ell^{2s-1}}\,\mathcal{L}^0[\mathcal{B}[\ser]](g^2/\ell^2)\nonumber\\
&=-\frac{8\pi^2}{g^2(2s)!}\left(\dop\right)_{2s-1}\sum_{\ell=1}^\infty\frac{1}{\ell^{2s-1}}\,\left(\frac{2\,\pi^{3/2}\,\ell}{g}\,e^{\frac{4\pi^2\ell^2}{g^2}}\,\left(1-\text{Erf}\left(\tfrac{2\pi\ell}{g}\right)\right)-1\right)\,.\label{SeriesBorelResumCorr}
\end{align}
This infinite sum of error-functions therefore provides an alternative representation of $\ztu{2s+1}$. Their relation is numerically shown for $s=1$ in Figure~\ref{Fig:ResummationProcedureI1}: the left panel shows a truncation of (\ref{SeriesZetaSingle}) at different orders
\begin{align}
\mathcal{I}_{s,1}(K):=-\frac{2\rho}{\Gamma(2s+1)}\sum_{k=1}^K (-1)^k\frac{\Gamma(2k+2s-1)}{\Gamma(k)\,2^k\,\rho^k}\,\zeta_{2k+2s-1}\,,\label{SeriesZetaSingleL}
\end{align}
while the right panel shows a truncation of (\ref{SeriesBorelResumCorr}) for different orders
\begin{align}
\mathcal{I}_{s,2}(L)&:=-\frac{8\pi^2}{g^2(2s)!}\left(\dop\right)_{2s-1}\sum_{\ell=1}^L\frac{1}{\ell^{2s-1}}\,\mathcal{L}^0[\mathcal{B}[\ser]](g^2/\ell^2)\,.\label{SeriesBorelResumCorrL}
\end{align}
As is visible from Figure~\ref{Fig:ResummationProcedureI1}, $\mathcal{I}_{s,2}(L)$ in (\ref{SeriesBorelResumCorrL}) converges to $\ztu{3}$ in the limit $L\to \infty$, while $\mathcal{I}_{s,1}(K)$ is an asymptotic series with vanishing radius of convergence: while initially converging to $\ztu{3}$, for large enough $K$, $\mathcal{I}_{s,1}(K)$ diverges for all values of $g$.

As a further remark, we illuminate the relation between (infinite combinations of) error functions with the Gauss-deformed zeta functions also in the opposite direction. To this end, we try to rewrite general combinations of the following form
\begin{align}
\mathcal{J}_s=\sum_{\ell=1}^\infty\frac{1}{\ell^{2s-1}}\,\ser\left(\tfrac{g^2}{\ell^2}\right)\sim\sum_{\ell=1}^\infty\frac{1}{\ell^{2s-1}}\,\mathcal{L}^0[\mathcal{B}[\ser]](g^2/\ell^2)\,,
\end{align}
as combinations of $\ztu{n}$. We represent this expression as the following (non-convergent) series
\begin{align}
\mathcal{J}_s=\sum_{k=1}^\infty \frac{(-1)^k (2k)!}{k! (16\pi^2)^k}\,g^{2k}\,\zeta_{2k+2s-1}\,,
\end{align} 
which we regularise by replacing the Riemann zeta functions as
\begin{align}
\mathcal{J}_{s,\delta}=2^{2s-2}\,\int_0^{\frac{1}{2\delta}}\frac{w^{2s-1}dw}{\sinh^2(w)}\,\sum_{k=1}^\infty \frac{(-1)^k (2k)!}{k! (16\pi^2)^k}\,\,\frac{(2gw)^{2k}}{\Gamma(2s+2k)}\,.
\end{align} 
After partial integration, this function can be written in the following form, up to terms which vanish for $\delta\to 0$
\begin{align}
\mathcal{J}_{s,\delta}=\int_0^{\frac{1}{\delta}}d^1[w]\,\left[e^{-\frac{g^2 w^2}{16\pi^2}}\,P_1(w)+\text{Erf}\left(\frac{gw}{4\pi}\right)\,P_2(w)+P_3(w)\right]\,,\label{InverseJEngineering}
\end{align}
which we have checked up to $s=15$ and where $P_{1,2,3}$ are polynomials in $w$, notably
\begin{align}
&P_2(w)=\frac{2^{2s-3}\pi^{2s-5/2}\,w}{g^{2s-3}\,\Gamma(s-1)}\,{}_1F_1\left(2-s,\tfrac{3}{2},-\tfrac{g^2 w^2}{16\pi^2}\right)\,,\nonumber\\
&P_3(w)=-\left(\frac{2\pi}{g}\right)^{2s-2}\,\sum_{a=0}^{s-1}\frac{(1/2)_{s-1-a}}{((-1/2)_{s-1-a})^2}\,\frac{1}{(2a)!}\,\left(\frac{gw}{2\pi}\right)^{2a}\,.
\end{align}
Notice, while the terms proportional to $P_1$ in (\ref{InverseJEngineering}) gives rise to a combination of \emph{finitely} many $\mathfrak{z}$-functions, the terms proportional to $P_2$ lead (through the error function) to an infinite series of $\mathfrak{z}$-functions, while $P_3$ gives rise to (combinations of finitely many) Riemann zeta functions. Therefore, $\mathcal{J}_{s,\delta}$ does not give rise to a simple combination of functions $\ztu{n}$ for finitely many values of $n$. However, the polynomials $P_{2,3}$ are such that
\begin{align}
\prod_{a=1}^{2s-3}\left(\dop+a\right)\,P_{2,3}(w)=0\,,
\end{align}
which explains why the in $\mathcal{Z}_\emptyset^{(2s),1}$ in (\ref{BorelResumSingleDepth}) are in fact represented by regularised finite combinations of $\mathfrak{z}$-functions in (\ref{SingleDepthZetaFunction}).
\subsection{Contribution of Depth 2}
We next consider the contribution of depth 2, \emph{i.e.} $\mathcal{Z}_\emptyset^{2}$ in (\ref{ZdepthDecomposition}). To extract the contribution of order $\mathcal{O}(\madj^{2s})$ we need to develop products of two coefficients $c_{2n}$ in (\ref{FormckCoef}). Extracting from the latter the contribution of order $\mathcal{O}(\madj^{2s})$ leads to the following infinite series
\begin{align}
\mathcal{Z}_{\emptyset}^{(2s),2}&=(-1)^s\sum_{k=2}^\infty (-1)^k \frac{\Gamma(k+3/2)}{\rho^{k+3/2}}\,2^{k+1}\sum_{\ell=1}^{k-1}\sum_{n=\ell}^{\ell+s-2}\frac{\zeta_{2n+1}}{n+1}\,\frac{\zeta_{2k+2s-2n-3}}{k+s-n-1}\,\big(^{\hspace{0.15cm}2(n+1)}_{2(n-\ell+1)}\big)\,\big(^{2(k+s-n-1)}_{\hspace{0.05cm}2(s+\ell-n-1)}\big)\,.\label{Depth2Series}
\end{align}
As in the case of depth 1 (see (\ref{Depth1Series})), the sum over $k$ has zero radius of convergence. In order to still give sense to (\ref{Depth2Series}), we shall as before discuss its Borel resummation and a regularisation.
\subsubsection{Borel Resummation}
Following the strategy of Section~\ref{Sect:BorelResumFirst}, we first consider a Borel resummation of (\ref{Depth2Series}). To this end, we first consider the finite sums over $\ell$ and $n$ and use the definition (\ref{DefZetaFunction}) of the Riemann zeta function
\begin{align}
\mathcal{C}_k&=\sum_{\ell=1}^{k-1}\sum_{n=\ell}^{\ell+s-2}\frac{\zeta_{2n+1}}{n+1}\,\frac{\zeta_{2k+2s-2n-3}}{k+s-n-1}\,\big(^{\hspace{0.15cm}2(n+1)}_{2(n-\ell+1)}\big)\,\big(^{2(k+s-n-1)}_{\hspace{0.05cm}2(s+\ell-n-1)}\big)=\sum_{n=0}^{s-2}\mathcal{C}_{k,n}\,.
\end{align}
with
\begin{align}
\mathcal{C}_{k,n}=\sum_{n_1,n_2=1}^\infty\sum_{\ell=1}^{k-1}\frac{\big(^{2(n+\ell+1)}_{\hspace{0.15cm}2(n+1)}\big)}{n_1^{2n+2\ell+1}(n+\ell+1)}\,\frac{\big(^{2(k+s-n-\ell-1)}_{\hspace{0.45cm}2(s-n-1)}\big)}{n_2^{2k+2s-2n-2\ell-3}(k+s-n-\ell-1)}\,.
\end{align}
For further convenience we, separate the sum over $n_1,n_2$ into a diagonal- and non-diagonal part
\begin{align}
\mathcal{C}_{k,n}&=\sum_{n_1=1}^\infty\frac{1}{n_1^{2k+2s-2}} \sum_{\ell=1}^{k-1}A(\ell)\,B(k-\ell)+\sum_{n_1>n_2\geq 1}\frac{1}{n_1^{2n+1}n_2^{2k+2s-2n-3}}\sum_{\ell=1}^{k-1}\left(\frac{n_2}{n_1}\right)^{2\ell}\,A(\ell)\,B(k-\ell)\,,\label{InterSumCkn}
\end{align}
where we have defined
\begin{align}
&A(\ell)=\frac{\big(^{2(n+\ell+1)}_{\hspace{0.15cm}2(n+1)}\big)}{(n+\ell+1)}\,,&&B(k-\ell)=\frac{\big(^{2(k-\ell+s-n-1)}_{\hspace{0.45cm}2(s-n-1)}\big)}{k-\ell+s-n-1}\,.
\end{align}
These are polynomials of their respective arguments which we expand with coefficients $a_i$ and $b_i$ respectively
\begin{align}
&A(u)=\sum_{i=0}^{2n+1}\alpha_i\,u^i\,,&&B(u)=\sum_{i=0}^{2(s-n)-3}\beta_i\, u^i\,. 
\end{align}
In order to write the sum over $\ell$ in (\ref{InterSumCkn}) also as a polynomial in $k$, we use the identity
\begin{align}
\sum_{\ell=1}^{k-1}\ell^m=\frac{B_{m+1}(k)-B_{m+1}(1)}{m+1}\,,
\end{align}
where $B_{m+1}(x)$ are the Bernoulli polynomials. We then have
\begin{align}
\sum_{\ell=1}^{k-1}\ell^i\,(k-\ell)^j=\sum_{\ell=1}^{k-1}\sum_{a=0}^j \big(^j_a\big)\,(-1)^a\,\ell^{a+i}\,k^{j-a}=\sum_{a=0}^j\big(^j_a\big)\,\,(-1)^a k^{j-a}\frac{B_{a+i+1}(k)-B_{a+i+1}(1)}{a+i+1}\,.
\end{align}
We furthermore use the following geometric series
\begin{align}
\sum_{\ell=1}^{k-1}x^\ell \ell^n=\left(x\,\frac{d}{dx}\right)^n\,\frac{x(1-x^{k-1})}{1-x}:=\frac{x}{(1-x)^{n+1}}\,\left(\sum_{j=0}^n\gamma_j(n)\,x^j-x^{k-1}\,\sum_{j=0}^n\,\delta_j(n,x)\,k^j\right)\,,
\end{align}
with $\gamma_j(n)$ constants and $\delta_j(n,x)$ polynomials in $x$ that depend on $n$. This allows to write the summations over $\ell$ in (\ref{InterSumCkn}) in the form of polynomials in $k$
\begin{align}
\mathcal{C}_{k,n}&=\sum_{n_1=1}^\infty\frac{1}{n_1^{2k+2s-2}} \sum_{i=0}^{2n+1}\sum_{j=0}^{2s-2n-3}\alpha_i\,\beta_j\,\sum_{a=0}^j\big(^j_a\big)\,\,(-1)^a k^{j-a}\frac{B_{a+i+1}(k)-B_{a+i+1}(1)}{a+i+1}\nonumber\\
&+\sum_{n_1>n_2\geq 1}\frac{1}{n_1^{2n+1}n_2^{2k+2s-2n-3}}\sum_{i=0}^{2n+1}\sum_{j=0}^{2s-2n-3}\alpha_i\,\beta_j\,\sum_{a=0}^j\,\big(^j_a\big)\,\,(-1)^a k^{j-a}\,\frac{n_2^2/n_1^2}{\left(1-\left(\tfrac{n_2}{n_1}\right)^{2a+2i+2}\right)}\nonumber\\
&\hspace{2cm}\times\sum_{c=0}^{a+i}\bigg(\gamma_c(a+i)\,\left(\tfrac{n_2}{n_1}\right)^{2c}-\left(\tfrac{n_2}{n_1}\right)^{2k-2}\delta_c(a+i,n_2/n_1)\,k^c\bigg)\,.
\end{align}
Substituting this result into (\ref{Depth2Series}) we can write $\mathcal{Z}_{\emptyset}^{(2s),2}$ in terms of the series (\ref{SeriesInitial}) 
\begin{align}
&\mathcal{Z}_{\emptyset}^{(2s),2}=\frac{(-1)^s\sqrt{\pi}}{\rho^{3/2}}(2\mathfrak{d}+1)\sum_{n=0}^{s-2}\sum_{i=0}^{2n+1}\sum_{j=0}^{2s-2n-3}\nonumber\\
&\times \bigg[\sum_{n_1=1}^\infty\frac{\alpha_i\,\beta_j}{n_1^{2s-2}} \,\sum_{a=0}^j\big(^j_a\big)\,\,(-1)^a \mathfrak{d}^{j-a}\frac{B_{a+i+1}(\mathfrak{d})-B_{a+i+1}(1)}{a+i+1}\ser\left(\tfrac{g^2}{n_1^2}\right)\nonumber\\
&\hspace{1cm}+\sum_{n_1>n_2\geq 1}\frac{1}{n_1^{2n+1}n_2^{2s-2n-3}}\sum_{i=0}^{2n+1}\sum_{j=0}^{2s-2n-3}\alpha_i\,\beta_j\,\sum_{a=0}^j\,\big(^j_a\big)\,\,(-1)^a \mathfrak{d}^{j-a}\,\frac{n_2^2/n_1^2}{\left(1-\left(\tfrac{n_2}{n_1}\right)^{2a+2i+2}\right)}\,\nonumber\\
&\hspace{2.7cm}\times\sum_{c=0}^{a+i}\bigg(\gamma_c(a+i)\,\left(\tfrac{n_2}{n_1}\right)^{2c}\,\ser\left(\tfrac{g^2}{n_2^2}\right)-\left(\tfrac{n_1}{n_2}\right)^{2}\delta_c(a+i,n_2/n_1)\,\mathfrak{d}^c\ser\,\left(\tfrac{g^2}{n_1^2}\right)\bigg)\bigg]\,.\label{Depth2SerRewrite}
\end{align}
where $\dop=g\,\frac{d}{dg}$ as in (\ref{ShiftZetaValue}). Resumming $\ser$ in the form (\ref{BorelResumS1}) therefore gives an infinite series of $\mathcal{L}^0[\mathcal{B}[\ser]]$ dressed with a non-trivial differential operator that are polynomials in $\dop$. While this shows that the basic building block are therefore again error-functions, they appear in much more complicated combinations than at depth 1.

\subsubsection{Regularisation}
\begin{figure}
\vspace{-0.25cm}
\begin{center}
\includegraphics[width = 7.5cm]{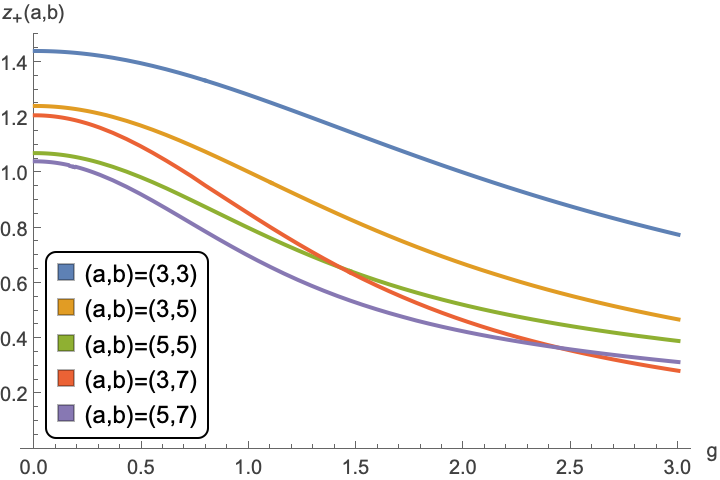}\hspace{1cm}\includegraphics[width = 7.5cm]{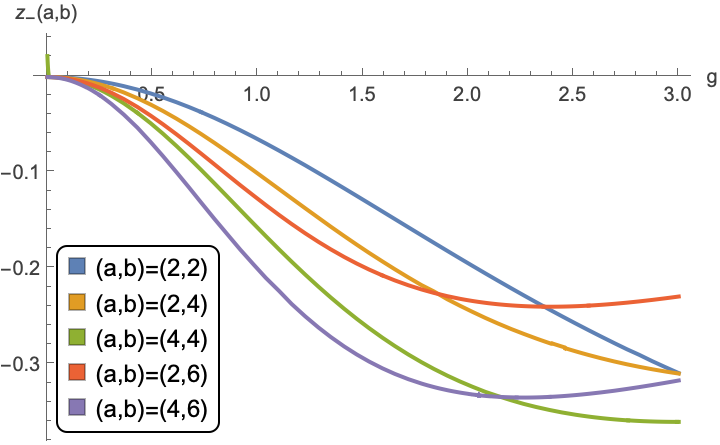}
\end{center}
\vspace{-0.5cm}
\caption{Numerical plots of $\ztua{a}{b}$ (left panel) and $\ztub{a}{b}$ (right panel) for certain $(a,b)$.}
\label{Fig:NumericalDoubleZeta}
\vspace{-0.25cm}
\end{figure}

We next consider a regularisation of (\ref{Depth2Series}): following the example of (\ref{Z2s1regularisation}), we define
\begin{align}
\mathcal{Z}_{\emptyset,\delta}^{(2s),2}&=(-1)^s\sum_{k=2}^\infty (-1)^k \frac{\Gamma(k+3/2)}{\rho^{k+3/2}}\,2^{k+1}\sum_{\ell=1}^{k-1}\sum_{n=\ell}^{\ell+s-2}\frac{1}{n+1}\,\frac{1}{k+s-n-1}\,\big(^{\hspace{0.15cm}2(n+1)}_{2(n-\ell+1)}\big)\,\big(^{2(k+s-n-1)}_{\hspace{0.05cm}2(s+\ell-n-1)}\big)\nonumber\\
&\hspace{1cm}\times \frac{2^{2n}}{\Gamma(2n+2)}\int_0^{\frac{1}{2\delta}}\frac{w^{2n+1}\,dw}{\sinh^2(w)}\,\frac{2^{2k+2s-2n-4}}{\Gamma(2k+2s-2n-2)}\int_0^{\frac{1}{2\delta}}\frac{v^{2k+2s-2n-3}\,dv}{\sinh^2(v)}\,.
\end{align}
The regularised integrals may be exchanged with the summation over $k$, leading to a (uniformely) converging series. Up to terms vanishing in the limit $\delta\to 0$ (indicated by a $\sim$), we may write $\mathcal{Z}_{\emptyset,\delta}^{(2s),2}$ in the following form 
\begin{align}
\mathcal{Z}_{\emptyset,\delta}^{(2s),2}&\sim\int_0^{\frac{1}{\delta}}d^2[w,v]\,\mathcal{P}_2(w,v;s):=\int_0^{\frac{1}{\delta}} \frac{dw}{e^{w}-1}\int_0^{\frac{1}{\delta}} \frac{dv}{e^v-1}\,\mathcal{P}_2(w,v;s)\,,\label{Zreg2s}
\end{align}
with 
\begin{align}
\mathcal{P}_2(w,v;s)&=e^{-\frac{(w+v)^2}{2\rho}}\,\mathbb{T}_+(w,v;s)+e^{-\frac{(w-v)^2}{2\rho}}\,\mathbb{T}_-(w,v;s)\nonumber\\
&\hspace{1cm}+e^{-\frac{w^2}{2\rho}}\,\mathbb{S}_1(w,v;s)+e^{-\frac{v^2}{2\rho}}\,\mathbb{S}_1(v,w;s)+\mathbb{S}_0(w,v;s)\,.\label{PFactForm}
\end{align}
Here, $\mathbb{T}_\pm$, $\mathbb{S}_1$ and $\mathbb{S}_0$ are polynomials in $w$ and $v$ that depend on $s$ and $\rho$. We can discuss the individual contributions to (\ref{Zreg2s}) separately in the limit $\delta\to 0$:
\begin{align}
\mathcal{Z}_{\emptyset}^{(2s),2}\sim \mathcal{Z}_{\emptyset,0}^{(2s),2}+\mathcal{Z}_{\emptyset,1}^{(2s),2}+\mathcal{Z}_{\emptyset,2}^{(2s),2}\,.
\end{align}
Here the individual contributions take the following form (which we have checked up to $s=20$)
\begin{itemize}
\item $\mathcal{Z}_{\emptyset,0}^{(2s),2}=\int_0^\infty d^2[w,v]\,\mathbb{S}_0(w,v;s)$ is determined by the polynomial
\begin{align}
\mathbb{S}_0(w,v;s)=\frac{(-1)^s\sqrt{\pi}}{\rho^{3/2}\Gamma(2s-3)}\sum_{n=0}^{s-2}\,\frac{w^{2 s-2 n-4}\,v^{2n}}{(n+1) (s-n-1)}\,\big(^{2 s-4}_{\,\,\,\,2 n}\big)\,.
\end{align}
In the limit $\delta\to 0$, the integrals over $w$ and $v$ in (\ref{Zreg2s}) lead to Riemann zeta functions upon using (\ref{MellinRepresentationZeta}), thus yielding the following contribution to (\ref{Zreg2s})
\begin{align}
\mathcal{Z}_{\emptyset,0}^{(2s),2}\sim_{\delta \to 0}\int_0^{\infty}d^2[w,v]\,\mathbb{S}_0(w,v;s)\sim_{\delta\to 0}\frac{(-1)^s\sqrt{\pi}}{\rho^{3/2}}\sum_{n=0}^{s-2}\frac{\zeta_{2n+1}\zeta_{2s-2n-3}}{(n+1)(s-n-1)}\,.\label{IntRepS0Term}
\end{align}
Notice that the terms for $n=0$ and $n=s-2$ lead to $\zeta_1$, which is divergent. Indeed, the integral over $v$ in (\ref{IntRepS0Term}) is divergent at $v=0$ for $n=0$ (and similarly the integral over $w$ diverges at $w=0$ for $n=s-2$). However, these divergences cancel similar singularities stemming from the remaining contributions in (\ref{PFactForm}), as we shall see below.
\item $\mathcal{Z}_{\emptyset,1}^{(2s),2}=\int_0^\infty d^2[w,v]\,\left(e^{-\frac{w^2}{2\rho}}\,\mathbb{S}_1(w,v;s)+e^{-\frac{v^2}{2\rho}}\,\mathbb{S}_1(v,w;s)\right)$: Using
\begin{align}
\mathbb{S}_1(w,v;s)=\frac{\sqrt{\pi}(-1)^{s+1}}{\rho^{7/2}}\,\sum_{n=0}^{s-2}\frac{w^{2s-2n-4}v^{2n}}{(n+1)(2n)!}\,\frac{2\left(w^4-2(s-n)w^2\rho+(2s-2n-3)\rho^2\right)}{\Gamma(2s-2n-1)}\,,\nonumber
\end{align}
we obtain in the limit $\delta\to 0$
{\allowdisplaybreaks
\begin{align}
&\mathcal{Z}_{\emptyset,1}^{(2s),2}\sim_{\delta \to 0} \int_0^{\infty}d^2[w,v]\,\left(e^{-\frac{w^2}{2\rho}}\,\mathbb{S}_1(w,v;s)+e^{-\frac{v^2}{2\rho}}\,\mathbb{S}_1(v,w;s)\right)\nonumber\\
&=\frac{2\sqrt{\pi}(-1)^{s-1}}{\rho^{7/2}}\sum_{n=0}^{s-2}\bigg(\frac{4(2s-2n-1)(s-n)\,\ztu{2s-2n+1}}{n+1}-\rho\,\frac{4(s-n)\,\ztu{2s-2n-1}}{n+1}\nonumber\\
&\hspace{4cm}+\rho^2\,\frac{\ztu{2s-2n-3}}{(n+1)(s-n-1)}\bigg)\,\zeta_{2n+1}\,,\label{InterS1Depth1}
\end{align}
where $\ztu{a}$ is defined in (\ref{SingleZetReg}). Similar to (\ref{IntRepS0Term}) also this expression contains divergent terms for $n=0$ and $n=s-2$, characterised either by $\zeta_1$ or $\ztu{1}$. Moreover, this result has strong resemblance with (\ref{SingleDepthZetaFunction}), the contribution to the partition function at depth 1. Indeed, combining (\ref{IntRepS0Term}) with (\ref{InterS1Depth1}) we can write
\begin{align}
\mathcal{Z}_{\emptyset,0}^{(2s),2}+\mathcal{Z}_{\emptyset,1}^{(2s),2}
&=\sum_{n=0}^{s-2}\zeta_{2n+1}\left(\frac{2(-1)^n}{n+1}\,\mathcal{Z}_{\emptyset}^{(2s-2n-2),1}-\frac{(-1)^s\sqrt{\pi}}{\rho^{3/2}}\frac{\zeta_{2s-2n-3}}{(n+1)(s-n-1)}\right)\,.\label{ProductDepth2}
\end{align}}
\item $\mathcal{Z}_{\emptyset,2}^{(2s),2}=\int_0^\infty d^2[w,v]\,\left(e^{-\frac{(w+v)^2}{2\rho}}\,\mathbb{T}_+(w,v;s)+e^{-\frac{(w-v)^2}{2\rho}}\,\mathbb{T}_-(w,v;s)\right)$: 
In order to treat the contributions related to $\mathbb{T}_\pm$ in (\ref{PFactForm}), we first consider terms proportional to $\mathbb{T}_++\mathbb{T}_-$. Concretely, we find
{\allowdisplaybreaks
\begin{align}
&\int_0^{\frac{1}{\delta}}d^2[w,v]\,\left(e^{-\frac{(w+v)^2}{2\rho}}+e^{-\frac{(w-v)^2}{2\rho}}\right)\,\frac{\mathbb{T}_++\mathbb{T}_-}{2}\nonumber\\
&\sim_{\delta\to 0}\frac{(-1)^s\sqrt{\pi}}{\rho^{9/2}}\bigg[\rho^3\sum_{n=0}^{s-2} \frac{\ztua{2s-2 n-3}{2 n+1}}{(s-n-1)(n+1)}\nonumber\\
&+4\rho^2\bigg(\sum_{n=0}^{s-1} \frac{ \left(s^2+1-2 n s+2 n (n+1)\right) \ztua{2s-2 n-1}{2 n+1}}{(n-s)(n+1)}+\frac{\ztua{2s-1}{1}+\ztua{1}{2s-1}}{s}\bigg)\nonumber\\
&+4\rho\bigg(\sum_{n=0}^{s} \frac{\left(6-4 n^2+4 n s+(s+1) (s (2 s+5))\right) \ztua{2s-2 n+1}{2 n+1}}{(s-n+1)(n+1)}\nonumber\\
&\hspace{2cm}-\frac{\ztua{3}{2 s-1}+\ztua{2s-1}{3}}{s}-6\,(s+1)\left(\ztua{2s+1}{1}+\ztua{1}{2s+1}\right)\bigg)\nonumber\\
&-32  \sum_{n=1}^{s}  \left(4 n^2-4 n (s+1)+(s+1) (2 s+1)\right) \ztua{2s-2 n+3}{2 n+1}\nonumber\\
&\hspace{2cm}+32  \left(\ztua{3}{2 s+1}+\ztua{2s+1}{3}\right)\bigg]\label{TplusContDepth2}
\end{align}}
where we have defined
\begin{align}
\mathfrak{z}_+(a,b)=\frac{1}{2\Gamma(a)\Gamma(b)}\int_0^\infty dt_1\int_0^\infty dt_2\frac{t_1^{a-1}t_2^{b-1}}{(e^{t_1}-1)(e^{t_2}-1)}\,\left(e^{-\frac{(t_1+t_2)^2}{2\rho}}+ e^{-\frac{(t_1-t_2)^2}{2\rho}}\right)\,.\label{DefZPlus}
\end{align}
We first remark that all sums in (\ref{TplusContDepth2}) are finite, \emph{i.e.} they represent combinations of finitely many functions of the form (\ref{DefZPlus}). Furthermore, all expressions are symmetric in the exchange of the arguments of the $\ztua{a}{b}$-functions (\emph{i.e.} for every $\ztua{a}{b}$ there is also a $\ztua{b}{a}$, which appears with the same coefficient). However, (\ref{TplusContDepth2}) also contains divergent contributions, since $\ztua{1}{s}$ and $\ztua{s}{1}$ diverge for any $s>1$, since the integrals over $w$ or $v$ diverge at $0$ in (\ref{DefZPlus}). However, these singular contributions are precisely cancelled by singular contributions in (\ref{ProductDepth2}) that we remarked earlier, so that the combination of (\ref{TplusContDepth2}) and (\ref{ProductDepth2}) is finite. Numerical plots of $\ztua{a}{b}$ for a number of $(a,b)$ are shown in the left panel of Figure~\ref{Fig:NumericalDoubleZeta}.

Finally it remains to consider the contribution proportional to $\mathbb{T}_+-\mathbb{T}_-$ in (\ref{PFactForm}). Concretely, we find 
{\allowdisplaybreaks
\begin{align}
&\int_0^{\frac{1}{\delta}}d^2[w,v]\,\left(e^{-\frac{(w+v)^2}{2\rho}}-e^{-\frac{(w-v)^2}{2\rho}}\right)\,\frac{\mathbb{T}_+-\mathbb{T}_-}{2}\nonumber\\
&\sim_{\delta\to 0}\frac{(-1)^s\sqrt{\pi}}{\rho^{9/2}}\,\bigg[\rho^2\sum_{n=0}^{s-2}\frac{(-8 n (n-s+2)+10 s-9)\, \ztub{2 s-2n-2}{2 n+2}}{(n+1) (n-s+1)}\nonumber\\
&+4\rho\,\bigg(\sum_{n=0}^{s-1}\frac{\left(5 s^2-4 n s+(2 n+1)^2+2 s^3\right)\ztub{2 s-2 n}{2 n+2}}{(n+1) (s-n)}-\frac{(3s+1)\left(\ztub{2}{2s}+\ztub{2s}{2}\right)}{s}\bigg)\nonumber\\
&+\sum_{n=0}^{s}\bigg(\frac{4 \ztub{2 s-2n+2}{2 n+2}}{(n+1) (n-s-1)} \big(32 n^4-64 n^3 s+8 n^2 \left(6 s^2-3 s-4\right)\nonumber\\
&\hspace{2.5cm}+8 n s ((3-2 s) s+4)+s^2 (4 s (s+1)+11)+17 s+6\bigg)\nonumber\\
&+\frac{12}{s}\,\left(\ztub{2s}{4}+\ztub{4}{2s}\right)+24(2s+1)\,\left(\ztub{2s+2}{2}+\ztub{2}{2s+2}\right)\bigg)\bigg]\,.\label{TminusContDepth2}
\end{align}}
where we have defined
\begin{align}
\mathfrak{z}_-(a,b)=\frac{1}{2\Gamma(a)\Gamma(b)}\int_0^\infty dt_1\int_0^\infty dt_2\frac{t_1^{a-1}t_2^{b-1}}{(e^{t_1}-1)(e^{t_2}-1)}\,\left(e^{-\frac{(t_1+t_2)^2}{2\rho}}- e^{-\frac{(t_1-t_2)^2}{2\rho}}\right)\,.\label{DefZMinus}
\end{align}
Similar to (\ref{TplusContDepth2}) also this expression contains only finite sums of $\ztub{a}{b}$ functions in a symmetric fashion. However, in contrast to (\ref{TplusContDepth2}), (\ref{TminusContDepth2}) contains no singular contributions. Numerical plots of $\ztub{a}{b}$ for a number of $(a,b)$ are shown in the right panel of Figure~\ref{Fig:NumericalDoubleZeta}.

The contributions (\ref{TplusContDepth2}) and (\ref{TminusContDepth2}) contain various different $\mathfrak{z}_\pm$ functions with different arguments. These functions satisfy the following relation
\begin{align}
-\rho\,\dop\,\mathfrak{z}_\pm(a,b)=a(a+1)\,\mathfrak{z}_\pm(a+2,b)+b(b+1)\,\mathfrak{z}_\pm(a,b+2)+2ab\,\mathfrak{z}_{\mp}(a+1,b+1)\,,\label{Zeta2Identity}
\end{align}
which generalises (\ref{SingleZetReg}) in the case $\ztu{a}$ (and where $\dop$ is also defined in (\ref{ShiftZetaValue})). Using the identity (\ref{Zeta2Identity}) iteratively, we have verified up to $s=18$ that the combined contribution of (\ref{TplusContDepth2}) and (\ref{TminusContDepth2}) simplifies to the following more compact expression
\begin{align}
&\mathcal{Z}_{\emptyset,2}^{(2s),2}=\int_0^\infty d^2[w,v]\,\left(e^{-\frac{(w+v)^2}{2\rho}}\,\mathbb{T}_+(w,v;s)+e^{-\frac{(w-v)^2}{2\rho}}\,\mathbb{T}_-(w,v;s)\right)\nonumber\\
&=\frac{(-1)^s\sqrt{\pi}}{\rho^{9/2}}\bigg[\rho^3(\dop+1)\left(\sum_{\ell=1}^{s-1}\frac{\mathcal{A}(\ell,s)}{\ell(s-\ell)}\left(\dop+\tfrac{1}{\ell(s-\ell)}\right)\ztua{2\ell - 1}{2s - 2\ell - 1}\right)\nonumber\\
&\hspace{0.2cm}+\rho^3(\dop+1)\dop\left(\sum_{\ell=1}^{s-2}\frac{\mathcal{B}(\ell,s)}{\ell(s-\ell)}\ztub{2 \ell}{2s - 2 \ell - 2}\right)+\rho^2\sum_{\ell=1}^{s-1}\frac{(\dop+1)(\dop+3)}{\ell(s-\ell)}\ztub{2\ell}{2s-2\ell}\bigg]\,.\label{Z22compact}
\end{align}
Here we have defined the following numerical prefactors 
\begin{align}
\mathcal{A}(\ell,s)&=-\sum_{k=1}^\ell\frac{2^{2k-1}\,\ell\,\left(\ell-k-\tfrac{1}{2}\right)_k\left(\ell-k+1\right)_{k-1}}{k(2k-1)!(2s-2k-1)}+\sum_{k=1}^{\ell-1}\frac{2^{2k}\left(\ell-k+\tfrac{3}{2}\right)_{k-1}\,\left(\ell-k\right)_{k+1}}{(2\ell-1)(2k-1)!(2s-2k)}\,,\nonumber\\
\mathcal{B}(\ell,s)&=-\sum_{k=1}^\ell\frac{2^{2k}\ell \,\left(\ell-k-\tfrac{1}{2}\right)_{k+1}\left(\ell-k+1\right)_{k-1}}{k(2k-1)!(2s-2k-1)(2 \ell - 2 k + 1)}+\sum_{k=1}^{\ell-1}\frac{2^{2k-1}\left(\ell-k+\tfrac{3}{2}\right)_{k-1}\,\left(\ell-k\right)_{k+1}}{(\ell - k + 1)(2k-1)!(2s-2k)}\,,\nonumber
\end{align}
where the Pochhammer symbol $(x)_n$ is defined in (\ref{DefPochhammer}). If we define the weight of the $\mathfrak{z}_\pm$ as the sum over its arguments, then the terms proportional to $\rho^3$ in the square bracket in (\ref{Z22compact}) both have weight $2s-2$, while the terms proportional to $\rho^2$ have weight $2s$.
\end{itemize}

\noindent
Combining the various contributions, we therefore obtain 
\begin{align}
&\mathcal{Z}_{\emptyset}^{(2s),2}=\sum_{n=0}^{s-2}\zeta_{2n+1}\left(\frac{2(-1)^n}{n+1}\,\mathcal{Z}_{\emptyset}^{(2s-2n-2),1}-\frac{(-1)^s\sqrt{\pi}}{\rho^{3/2}}\frac{\zeta_{2s-2n-3}}{(n+1)(s-n-1)}\right)\nonumber\\
&\hspace{0.2cm}+\frac{(-1)^s\sqrt{\pi}}{\rho^{9/2}}\bigg[\rho^3(\dop+1)\left(\sum_{\ell=1}^{s-1}\frac{\mathcal{A}(\ell,s)}{\ell(s-\ell)}\left(\dop+\tfrac{1}{\ell(s-\ell)}\right)\ztua{2\ell - 1}{2s - 2\ell - 1}\right)\nonumber\\
&\hspace{0.5cm}+\rho^3(\dop+1)\dop\left(\sum_{\ell=1}^{s-2}\frac{\mathcal{B}(\ell,s)}{\ell(s-\ell)}\ztub{2 \ell}{2s - 2 \ell - 2}\right)+\rho^2\sum_{\ell=1}^{s-1}\frac{(\dop+1)(\dop+3)}{\ell(s-\ell)}\ztub{2\ell}{2s-2\ell}\bigg]\,.\label{FullDepth2Contr}
\end{align}
The concrete exemple $s=2$ is discussed in more detail in Section~\ref{Sect:OrderEps4} along with graphical visualisations of $\mathcal{Z}_{\emptyset}^{(2s),2}$ as a function of $g$.
\subsection{Contribution of Depth 3}
\subsubsection{Series Representation}
In order to calculate the contribution with three factors of $c_{2k}$ in (\ref{FormckCoef}), we use the definition (\ref{DefXiFun}) and compute
\begin{align}
c_{2i}\,c_{2j}\,c_{2m}&=(-1)^{i+j+m}\,2^{3+i+j+m}\,\sum_{a=i}^\infty\sum_{b=j}^\infty\sum_{c=m}^\infty\frac{\zeta_{2a+1}}{a+1}\,\frac{\zeta_{2b+1}}{b+1}\,\frac{\zeta_{2c+1}}{c+1}\,\big(^{\hspace{0.2cm}2a+2}_{2a+2-2i}\big)\,\big(^{\hspace{0.2cm}2b+2}_{2b+2-2j}\big)\,\big(^{\hspace{0.2cm}2c+2}_{2c+2-2m}\big)\nonumber\\
&\hspace{1cm}\times (-1)^{a-i+b-j+c-m+3}\,\madj^{2a-2i+2b-2j+2c-2m+6}\,. 
\end{align}
To extract the term of order $\madj^{2s}$ we need to impose
\begin{align}
&a+b+c=s+i+j+m-3&&\text{such that} &&c=s+i+j+m-a-b-3
\end{align}
which allows to remove one of the sums
\begin{align}
c_{2i}\,c_{2j}\,c_{2m}&=(-1)^{s+i+j+m}\,\madj^{2s}\,2^{6+i+j+m}\,\sum_{{a=i}\atop{b=j}}^{s+i+j-a-b-3\geq 0}\zeta_{2a+1}\,\zeta_{2b+1}\,\zeta_{2(s+i+j+m-a-b)-5}\nonumber\\
&\hspace{0.5cm}\times \frac{(2a+1)!}{(2a+2-2i)!(2i)!}\,\frac{(2b+1)!}{(2b+2-2j)!(2j)!}\,\frac{(2(s+i+j+m-a-b)-5)!}{(2(s+i+j-a-b-2))!(2m)!}\,. 
\end{align}
This allows us to compute the contribution of order $\madj^{2s}$
\begin{align}
\mathcal{Z}_{\emptyset}^{(2s),3}&=\frac{(-1)^s\sqrt{\pi}}{6\rho^{9/2}}\sum_{a=0}^{s-3}\sum_{k=4}^\infty\frac{(-1)^k 2^{6+k}\Gamma(k+3/2)}{\rho^{k-3}}\sum_{\ell=1}^{k-2}\,\frac{\zeta_{2a+2\ell+1}\,(2a+2\ell+1)!}{(2a+2)!(2\ell)!}\nonumber\\
&\hspace{0.2cm}\times\sum_{n=1}^{k-\ell-1} \sum_{b=n}^{s+n-a-3} \frac{\zeta_{2b+1}(2b+1)!}{(2b+2-2n)!(2n)!}\,\frac{\zeta_{2(s+k-\ell-a-b)-5}(2(s+k-\ell-a-b)-5)!}{(2(s+n-a-b-2))!(2k-2\ell-2n)!}\,.\label{Depth3Series}
\end{align}
\subsubsection{Regularisation}
\begin{figure}
\vspace{-0.25cm}
\begin{center}
\includegraphics[width = 7.5cm]{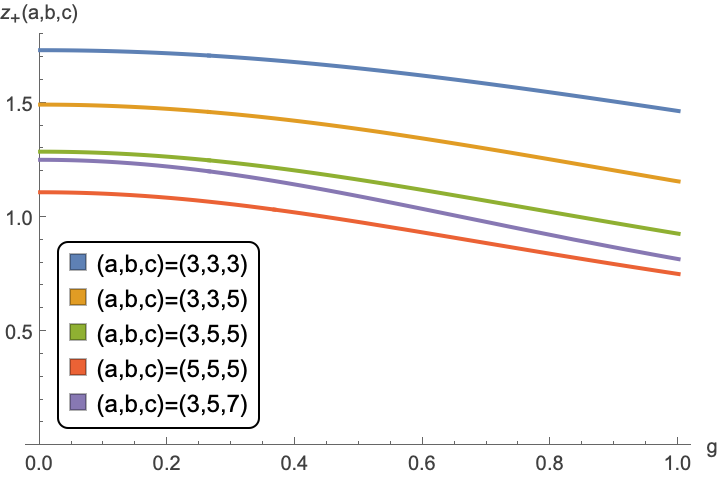}\hspace{1cm}\includegraphics[width = 7.5cm]{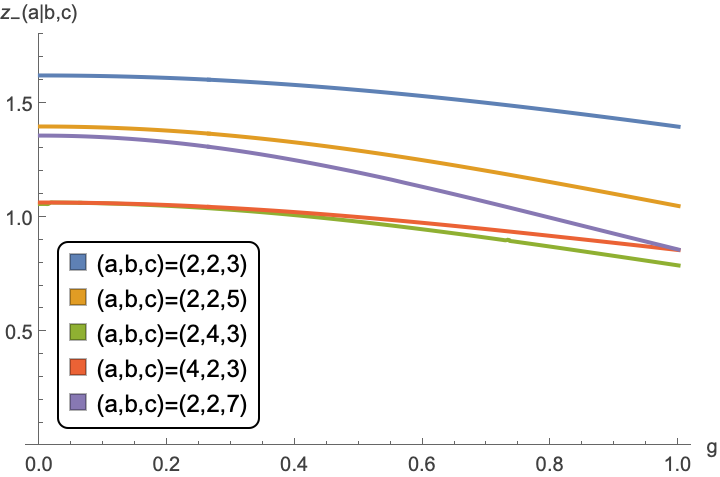}
\end{center}
\vspace{-0.5cm}
\caption{Numerical plots of $\mathfrak{z}_+(a,b,c)$ (left panel) and $\mathfrak{z}_-(a|b,c)$ (right panel) for certain $(a,b,c)$.}
\label{Fig:NumericalTrippleZeta}
\vspace{-0.25cm}
\end{figure}

As at depth 1 and 2 before, the summation over $k$ in (\ref{Depth3Series}) has zero radius of convergence. Given the complexity of the Borel resummation, here we shall only discuss the regularisation of (\ref{Depth3Series}), following the same prescription as in Section~\ref{Sect:Depth1Reg}. Indeed, to this end, we define
\begin{align}
\mathcal{Z}_{\emptyset,\delta}^{(2s),3}&=\frac{(-1)^s\sqrt{\pi}}{6\rho^{9/2}}\sum_{a=0}^{s-3}\sum_{k=4}^\infty\frac{(-1)^k 2^{6+k}\Gamma(k+3/2)}{\rho^{k-3}}\sum_{\ell=1}^{k-2}\,\frac{(2a+2\ell+1)!}{(2a+2)!(2\ell)!}\int_0^{\frac{1}{2\delta}}\frac{dw\,w^{2a+2\ell+1}}{\Gamma(2a+2\ell+2)}\nonumber\\
&\hspace{0.2cm}\times\sum_{n=1}^{k-\ell-1} \sum_{b=n}^{s+n-a-3} \frac{(2b+1)!}{(2b+2-2n)!(2n)!}\,\frac{(2(s+k-\ell-a-b)-5)!}{(2(s+n-a-b-2))!(2k-2\ell-2n)!}\nonumber\\
&\hspace{0.2cm}\times 2^{2s+2k-2\ell-2a-6}\int_0^{\frac{1}{2\delta}}\frac{du\,u^{2b+1}}{\Gamma(2b+2)}\int_0^{\frac{1}{2\delta}}\frac{dv v^{2s+2k-2\ell-2a-2b-5}}{\Gamma(2s+2k-2\ell-2a-2b-4)}\,,
\end{align}
with $\delta\ll 1$. The additional $\Gamma$-factors render the summations (notably over $k$) uniformly convergent, such that we may write (up to terms that vanish for $\delta\to 0$)
\begin{align}
\mathcal{Z}_{\emptyset,\delta}^{(2s),3}\sim\int_0^{\frac{1}{\delta}}d^3[w,v,u]\,\mathcal{P}_3(w,v,u;s):=\int_0^{\frac{1}{\delta}}\frac{dw}{e^w-1}\int_0^{\frac{1}{\delta}}\frac{dv}{e^v-1}\int_0^{\frac{1}{\delta}}\frac{du}{e^u-1}\,\mathcal{P}_3(w,v,u;s)\,.\label{Depth3PForm}
\end{align}
The integrand $\mathcal{P}_3(w,v,u;s)$ can be decomposed in the following form
\begin{align}
&\mathcal{P}_3(w,v,u;s)=e^{-\frac{(w+v+u)^2}{2\rho}}\,\mathbb{U}_+(w,v,u;s)+e^{-\frac{(-w+v+u)^2}{2\rho}}\,\mathbb{U}_-(w,v,u;s)\nonumber\\
&\hspace{0.5cm}+e^{-\frac{(w-v+u)^2}{2\rho}}\,\mathbb{U}_-(v,w,u;s)+e^{-\frac{(w+v-u)^2}{2\rho}}\,\mathbb{U}_-(u,v,w;s)\nonumber\\
&\hspace{0.5cm}+e^{-\frac{(w+v)^2}{2\rho}}\,\mathbb{T}_+(w,v,u;s)+e^{-\frac{(w+u)^2}{2\rho}}\,\mathbb{T}_+(u,v,w;s)+e^{-\frac{(v+u)^2}{2\rho}}\,\mathbb{T}_+(w,u,v;s)\nonumber\\
&\hspace{0.5cm}+e^{-\frac{(w-v)^2}{2\rho}}\,\mathbb{T}_-(w,v,u;s)+e^{-\frac{(w-u)^2}{2\rho}}\,\mathbb{T}_-(u,v,w;s)+e^{-\frac{(v-u)^2}{2\rho}}\,\mathbb{T}_-(w,u,v;s)\nonumber\\
&\hspace{0.5cm}+e^{-\frac{w^2}{2\rho}}\,\mathbb{S}_1(w,v,u;s)+e^{-\frac{v^2}{2\rho}}\,\mathbb{S}_1(v,w,u;s)+e^{-\frac{u^2}{2\rho}}\,\mathbb{S}_1(u,v,w;s)+\mathbb{S}_0(w,v,u;s)\,,\label{Depth3Exponentials}
\end{align}
with $\mathbb{U}_\pm$, $\mathbb{T}_\pm$, $\mathbb{S}_1$ and $\mathbb{S}_0$ polynomials in $(w,v,u)$ that depend on $s$ and $\rho$. In the limit $\delta\to 0$, we can thus write
\begin{align}
\mathcal{Z}_{\emptyset}^{(2s),3}\sim \mathcal{Z}_{\emptyset,0}^{(2s),3}+\mathcal{Z}_{\emptyset,1}^{(2s),3}+\mathcal{Z}_{\emptyset,2}^{(2s),3}+\mathcal{Z}_{\emptyset,3}^{(2s),3}\,.\label{SplittingDepth3}
\end{align}
Here the individual contributions take the following form (which we have checked up to $s=12$):
\begin{itemize}
\item $\mathcal{Z}_{\emptyset,0}^{(2s),3}=\int_0^\infty d^3[w,v,u]\,\mathbb{S}_0(w,v,u;s)$: The terms containing no exponential factor in (\ref{Depth3Exponentials}) give rise to (trilinear) combinations of Riemann zeta functions
\begin{align}
\mathcal{Z}_{\emptyset,0}^{(2s),3}=\frac{2\sqrt{\pi}(-1)^s}{3\rho^{3/2}}\,\sum_{{i,j,k\geq 1}\atop{i+j+k=s}}\left(\frac{840}{\rho^3}\left(i+\tfrac{1}{2}\right)\left(j+\tfrac{1}{2}\right)\left(k+\tfrac{1}{2}\right)\zeta_{2i+1}\zeta_{2j+1}\zeta_{2k+1}-\frac{\zeta_{2i-1}\zeta_{2j-1}\zeta_{2k-1}}{ijk}\right)\,.\label{ConstTermDepth3}
\end{align}
Notice that the second term in the bracket contains divergent $\zeta_1$ (similar to (\ref{IntRepS0Term}) in the case of depth 2), which, however, cancel similar divergences stemming from the remaining terms in (\ref{Depth3Exponentials}), as we shall discuss below.
\item $\mathcal{Z}_{\emptyset,1}^{(2s),3}=\int_0^\infty d^3[w,v,u]\,\left(e^{-\frac{w^2}{2\rho}}\,\mathbb{S}_1(w,v,u;s)+e^{-\frac{v^2}{2\rho}}\,\mathbb{S}_1(v,w,u;s)+e^{-\frac{u^2}{2\rho}}\,\mathbb{S}_1(u,v,w;s)\right)$

\noindent
The terms proportional to $\mathbb{S}_1$ can be partly rewritten using the contributions at depth 1 in (\ref{SingleDepthZetaFunction})
\begin{align}
\mathcal{Z}_{\emptyset,1}^{(2s),3}&= \sum_{n=1}^{s-2}\left(\mathcal{Z}_\emptyset^{(2n),1}\,\sum_{k=0}^{s-n-2}\frac{2(-1)^n\zeta_{2k+1}\zeta_{2s-2n-2k-3}}{(k+1)(s-n-k-1)}\right)+\frac{2\sqrt{\pi}(-1)^s}{\rho^{3/2}}\sum_{{i,j,k\geq 1}\atop{i+j+k=s}}\frac{\zeta_{2i-1}\zeta_{2j-1}\zeta_{2k-1}}{ijk}\,.\label{Depth3InterS2Contri}
\end{align}
Notice that the last term contains singular contributions in the form of $\zeta_1$, which cancel similar divergent terms in (\ref{ConstTermDepth3}) and the remaining contributions to (\ref{SplittingDepth3}).

\item $\mathcal{Z}_{\emptyset,2}^{(2s),3}$: We have verified up to order $s=12$ that the terms involving $\mathbb{T}_\pm$ in (\ref{Depth3PForm}) can be expressed in terms of $\mathcal{Z}_{\emptyset,2}^{(2s),2}$ in (\ref{Z22compact}), concretely
\begin{align}
\mathcal{Z}_{\emptyset,2}^{(2s),3}&=\int_0^\infty d^3[w,v,u]\bigg(e^{-\frac{(w+v)^2}{2\rho}}\,\mathbb{T}_+(w,v,u;s)+e^{-\frac{(w+u)^2}{2\rho}}\,\mathbb{T}_+(u,v,w;s)\nonumber\\
&\hspace{3.5cm}+e^{-\frac{(v+u)^2}{2\rho}}\,\mathbb{T}_+(w,u,v;s)+e^{-\frac{(w-v)^2}{2\rho}}\,\mathbb{T}_-(w,v,u;s)\nonumber\\
&\hspace{3.5cm}+e^{-\frac{(w-u)^2}{2\rho}}\,\mathbb{T}_-(u,v,w;s)+e^{-\frac{(v-u)^2}{2\rho}}\,\mathbb{T}_-(w,u,v;s)\bigg)\nonumber\\
&=\sum_{n=0}^{s-2}(-1)^n \frac{\zeta_{2n+1}}{n+1}\,\mathcal{Z}_{\emptyset,2}^{(2s-2n-2),2}\,.\label{Depth3TinterContri}
\end{align}
which corresponds to the same type of recursive recursive structure as in (\ref{ProductDepth2}) at depth~2.

\item $\mathcal{Z}_{\emptyset,3}^{(2s),3}$: developing all contributions (notably stemming from $\mathbb{U}_\pm$)
\begin{align}
\mathcal{Z}_{\emptyset,3}^{(2s),3}=\int_0^\infty d^3[w,v,u]\bigg(&e^{-\frac{(w+v+u)^2}{2\rho}}\,\mathbb{U}_+(w,v,u;s)+e^{-\frac{(-w+v+u)^2}{2\rho}}\,\mathbb{U}_-(w,v,u;s)\nonumber\\
&+e^{-\frac{(w-v+u)^2}{2\rho}}\,\mathbb{U}_-(v,w,u;s)+e^{-\frac{(w+v-u)^2}{2\rho}}\,\mathbb{U}_-(u,v,w;s)\bigg)\,,
\end{align}
is too complicated in full generality and beyond the scope of this work. We therefore limit ourselves to providing the generic building blocks in the leading examples. Indeed, we have checked up to $s=12$ that all $\mathcal{Z}_{\emptyset,3}^{(2s),3}$ can be written as finite contributions of the following two functions
\begin{align}
&\mathfrak{z}_+(a,b,c)=\int_0^\infty \frac{d^3[t_1,t_2,t_3]\,t_{(1}^{a-1}t_2^{b-1}t_{3)}^{c-1}}{4\Gamma(a)\Gamma(b)\Gamma(c)}\,\bigg(e^{-\frac{(t_1+t_2+t_3)^2}{2\rho}}+e^{-\frac{(-t_1+t_2+t_3)^2}{2\rho}}\nonumber\\
&\hspace{7.5cm}+e^{-\frac{(t_1-t_2+t_3)^2}{2\rho}}+e^{-\frac{(t_1+t_2-t_3)^2}{2\rho}}\bigg)\,,\label{DefZ3Pos}\\
&\mathfrak{z}_-(a|b,c)=\int_0^\infty \frac{d^3[t_1,t_2,t_3] t_{1}^{a-1}t_{(2}^{b-1}t_{3)}^{c-1}}{4\Gamma(a)\Gamma(b)\Gamma(c)}\,\left(e^{-\frac{(t_1+t_2+t_3)^2}{2\rho}}-e^{-\frac{(-t_1+t_2+t_3)^2}{2\rho}}\right)\,,\label{DefZ3Neg}
\end{align}
where the notation $(\ldots)$ means weighted symmetrisation, \emph{e.g.}
\begin{align}
t_{(1}^{a-1}t_2^{b-1}t_{3)}^{c-1}:=\frac{1}{3!}\,(&t_{1}^{a-1}t_2^{b-1}t_{3}^{c-1}+t_{1}^{a-1}t_3^{b-1}t_{2}^{c-1}+t_{2}^{a-1}t_1^{b-1}t_{3}^{c-1}+t_{2}^{a-1}t_3^{b-1}t_{1}^{c-1}\nonumber\\
&+t_{3}^{a-1}t_2^{b-1}t_{1}^{c-1}+t_{3}^{a-1}t_1^{b-1}t_{2}^{c-1})\,.
\end{align}
Graphical plots of $\mathfrak{z}_+(a,b,c)$ and $\mathfrak{z}_-(a|b,c)$ for some values of $(a,b,c)$ are shown in Figure~\ref{Fig:NumericalTrippleZeta}. Furthermore, as example, we find to leading order in $s=3$ 
\begin{align}
\mathcal{Z}_{\emptyset,3}^{(6),3}&=\frac{\sqrt{\pi}}{16\,\rho^{11/2}}\big[\rho^4\,\mathfrak{z}_+(1,1,1)-\rho^3\,\left(24\,\mathfrak{z}_+(1,1,3)+11\,\mathfrak{z}_-(2|1,2)\right)\nonumber\\
&+\rho^2\,\left(72\,\mathfrak{z}_+(1,1,5)+72\,\mathfrak{z}_+(1,3,3)+60\,\mathfrak{z}_-(2|1,4)+114\,\mathfrak{z}_-(2|2,3)+60\, \mathfrak{z}_-(4|1,2)\right)\nonumber\\
&-\rho\,\big(384\,\left(3\, \mathfrak{z}_+(1, 3, 5) + 2\, \mathfrak{z}_+(3, 3, 3)\right)+24\, (5\,\mathfrak{z}_-(2|1,6)+19\,\mathfrak{z}_-(2|2,5)\nonumber\\
&\hspace{1cm}+26\,\mathfrak{z}_-(2|3,4)+9\,\mathfrak{z}_-(4|1,4)+26\,\mathfrak{z}_-(4|2,3)+5\,\mathfrak{z}_-(6|1,2))\big)\nonumber\\
&+5760\, \mathfrak{z}_+(3, 3, 5)+240 \big(3\,\mathfrak{z}_-(2|2,7)+5\, \mathfrak{z}_-(2|3,6)+6\,\mathfrak{z}_-(2|4,5)+6\,\mathfrak{z}_-(4|2,5)\nonumber\\
&\hspace{1cm}+9\, \mathfrak{z}_-(4|3,4)+5\,\mathfrak{z}_- (6|2,3)\big)\big]\,.\label{ExplDepth3S1}
\end{align}
As a general feature, we remark that the $\mathfrak{z}_+(a,b,c)$ contributing to $\mathcal{Z}_{0,3}^{(2s),3}$ only carry odd arguments (\emph{i.e.} $a,b,c\in\mathbb{Z}_{\geq 0,\text{odd}}$), while among the arguments of $\mathfrak{z}_-(a|b,c)$, $a\in\mathbb{Z}_{\geq 0,\text{even}}$ is even and the sum $b+c\in\mathbb{Z}_{\geq 0,\text{odd}}$ is odd. Moreover, calling the weight of a $\mathfrak{z}_\pm$-function the sum of its arguments, the terms of order $\rho^\ell$ in (\ref{ExplDepth3S1}) all have weight $11-2\ell$.

\end{itemize}
 We remark that any $\mathfrak{z}$ carrying at least one argument $1$ is a priori divergent, since the respective integral in (\ref{SingleZetReg}), (\ref{TplusContDepth2}), (\ref{TminusContDepth2}), (\ref{DefZ3Pos}) or (\ref{DefZ3Neg}) diverges at $0$. The later should be thought of as being regularised by a parameter $\epsilon\ll 1$ in the same fashion as in (\ref{RegularisationEpsilon}) (and $\zeta_1\sim-\log\epsilon $ as in (\ref{RegularisationEpsilon})). However, combining all contributions as in (\ref{SplittingDepth3}) yields a finite $\mathcal{Z}_\emptyset^{(2s),3}$ in the limit $\epsilon\to 0$.

\subsection{Observations and Generalisations}
Before discussing in more detail the leading mass contributions $\mathcal{O}(\madj^2)$ and $\mathcal{O}(\madj^4)$ in the following Section, we summarise and  highlight a few general observations from the computations at different depths. We recall that the perturbative part of the partition function (\ref{Z00sumD}) can be written as a sum over contributions of different depth $\mathcal{Z}_{\emptyset}^r$ (see (\ref{DecompDepth})), where the expressions to  leading $r$ are shown in (\ref{ZdepthDecomposition}). This decomposition is not linked to a physical parameter of the system, but rather counts the number of coefficients $c_{2k}$ in (\ref{Z00sumD}), each of which given as a series of Riemann zeta functions. Extracting from the contribution of given depth $r$, the contribution at mass order $\mathcal{O}(\madj^{2s})$ (which is called $\mathcal{Z}_\emptyset^{(2s),r}$) leads to an asymptotic series of vanishing radius of convergence (see (\ref{Depth1Series}), (\ref{Depth2Series}) and (\ref{Depth3Series}) for depth $r=1,2,3$ respectively). In order to make sense of the latter, we have followed two strategies:
\subsubsection{Borel resummation} In all cases discussed, the asymptotic series can be written in terms of a single building block $\ser$ (see (\ref{SingleDepthGenBuildingBlock}) for $r=1$ and (\ref{Depth2SerRewrite}) for $r=2$), which is discussed in detail in Appendix~\ref{App:BorelResumSer}. More concretely, the contributions take the form of nested converging sums of $\ser$ dressed with (finite) combinations of differential operators of the form $(\dop+a)$ (with $\dop=g\tfrac{d}{dg}$ and $a\in\mathbb{R}$). $\mathcal{S}$ itself is an asymptotic series, whose Borel transform $\mathcal{B}[\ser](s)$ (see (\ref{BorelTrafoSer})) has no singular points for $\text{Re}(s)>0$. With the help of a directional Laplace-transform, we can therefore Borel resum $\ser$ in a unique way (see (\ref{BorelResumS1})), which yields notably an error-function. As remarked in \cite{Aniceto:2026pcz} in a very similar context, this fact is contingent on $g\in\mathbb{R}$ (\emph{i.e.} the coupling constant being physical), as otherwise singularities of $\Delta_{\text{1-loop}}$ in (\ref{OneLoopDet}) (located in an unphysical region of the parameter space) would lead to singularities in the Borel plane that give rise to a Stokes phenomenon.

Replacing $\ser$ by its Borel resummation $\mathcal{L}^0[\mathcal{B}[\ser]]$ thus allows to represent $\mathcal{Z}_\emptyset^{(2s),r}$ as (infinite) series of error functions, acted upon by specific differential operators as mentioned above. In contrast to the starting asymptotic series, however, these infinite series are generally convergent: we shall discuss the case $\mathcal{Z}_\emptyset^{(2)}=\mathcal{Z}_\emptyset^{(2),1}$ in more detail in Section~\ref{Sect:LeadingMassDeformation} below. Presentations of the partition function in this form (also in related contexts) have already previously been noted in the literature (see \emph{e.g.} \cite{Dorigoni:2024dhy,Aniceto:2014hoa,Aniceto:2026pcz}). However, in particular the example of $r=2$ (\ref{Depth2SerRewrite}) shows that the resulting expressions are still quite involved and cumbersome: at generic depth $r$, we expect the result to contain $r$ nested sums (\emph{i.e.} over parameters $n_1>n_2>\ldots>n_r\geq 1$) of (combinations of) $\ser(g^2/n_i^2)$ dressed with polynomials of the operator $\dop$. While we expect these different contributions to contain internal patterns and non-trivial (recursive) relations, we have not pursued these in the current work (\emph{e.g.} by studying in more detail the contribution at $r=3$). Instead, we have focused our efforts on the regularisation of the initial asymptotic series.

\subsubsection{Regularisation} Replacing the Riemann zeta functions by regularised Mellin integrals renders the asymptotic series  ((\ref{Depth1Series}) at depth $r=1$, (\ref{Depth2Series}) at depth $r=2$ and (\ref{Depth3Series}) at depth $r=3$) convergent, which can schematically be written in the form
\begin{align}
\mathcal{Z}_\emptyset^{(2s),r}= \sum_{k=0}^r\mathcal{Z}_{\emptyset,k}^{(2s),r}\,,\label{DepthSubDivision}
\end{align}
with the contributions
\begin{align}
\mathcal{Z}_{\emptyset,k}^{(2s),r}\sim \int_0^{\infty}d^r[\underline{t}]\,\sum_{{S\subset\{1,\ldots,r\}}\atop{k=|S|}}\,\sum_{\underline{\xi}\in \{\pm 1\}^k/\{\pm 1\}}\text{exp}\left(-\frac{1}{2\rho}\left(\sum_{a\in S}\xi_a t_a\right)^2\right)\,\mathbb{P}_{s,S}(\underline{t},\underline{\xi})\,,\label{GenFormZetaContributions}
\end{align}
with the integral measure $d^n[\underline{t}]:=\prod_{i=1}^r\frac{dt_i}{e^{t_i}-1}$ (and $\underline{t}=(t_1,\ldots,t_r)$) and $k$ denoting the cardinality of the set $S$. Furthermore, the summation over $\underline{\xi}$ is over $2^{k-1}$ many $k$-tuples with entries $\pm 1$\footnote{The $k$-tuples are labelled by $a\in S$, such that $\xi_a\in\{\pm 1\}$ $\forall a \in S$.} that are distinct modulo an overall sign, \emph{i.e.} $\underline{\xi}\sim-\underline{\xi} $: indeed, for given $S$ (with $|S|=k$) there are $2^{k-1}$ distinct combinations of exponentials $\text{exp}\left(-\frac{1}{2\rho}\left(\sum_{a\in S}\xi_a t_a\right)^2\right)$. The $\mathbb{P}_{s,S}(\underline{t},\underline{\xi})$ are polynomials in the variables $t_{1,\ldots,n}$ that are specific to $(\xi_1,\ldots,\xi_k)$ and $s$ (and also implicitly depend on $\rho$). The integrals in (\ref{GenFormZetaContributions}) lead to linear combinations of Gauss-deformed zeta functions with $k$ arguments (of the type introduced in (\ref{SingleZetReg}), (\ref{DefZPlus}), (\ref{DefZMinus}), (\ref{DefZ3Pos}) or (\ref{DefZ3Neg})) and $n-k$ Riemann zeta functions. Using symmetries of the $\mathbb{P}_{s,S}(\underline{t};\underline{\xi})$ (with $|S|=k$), we can define the following Gauss-deformed zeta functions
\begin{align}
\mathfrak{z}_{+}(a_1&,\ldots,a_k)=\int_0^\infty d^k[w_1,\ldots,w_k]\,\frac{w_{(1}^{a_1-1}\ldots w_{k)}^{a_k-1}}{2^{k-1}\prod_{i=1}^k\Gamma(a_i)}\sum_{\underline{\xi}\in \{\pm 1\}^k/\{\pm 1\}}\text{exp}\left(-\frac{1}{2\rho}\left(\sum_{i=1}^k\xi_i w_i\right)^2\right)\,,\label{GenZetaGenDepthPlus}
\end{align} 
which generalises the definitions (\ref{SingleZetReg}), (\ref{DefZPlus}) and (\ref{DefZ3Pos}), as well as 
\begin{align}
\mathfrak{z}_{-}(a_1&,\ldots,a_\ell|b_1\ldots,b_{k-\ell})=\int_0^\infty d^k[w_1,\ldots,w_\ell,v_1,\ldots,v_{k-\ell}]\,\frac{w_{(1}^{a_1-1}\ldots w_{\ell)}^{a_\ell-1}\,v_{(1}^{b_1-1}\ldots v_{k-\ell)}^{b_{k-\ell}-1}}{2^{k-1}\prod_{i=1}^\ell\Gamma(a_i)\prod_{j=1}^{k-\ell}\Gamma(b_j)}\nonumber\\
&\times \left[\text{exp}\left(-\frac{1}{2\rho}\left(\sum_{i=1}^{\ell}w_i+\sum_{j=1}^{k-\ell}v_j\right)^2\right)-\text{exp}\left(-\frac{1}{2\rho}\left(\sum_{i=1}^{\ell}w_i-\sum_{j=1}^{k-\ell}v_j\right)^2\right)\right]\,,\label{GenZetaGenDepthMinus}
\end{align} 
which generalise (\ref{DefZMinus}) and (\ref{DefZ3Pos}). The notation $(\ldots)$ denotes weighted symmetrisation of all indices, \emph{i.e.}
\begin{align}
w_{(1}^{a_1-1}\ldots w_{\ell)}^{a_\ell-1}=\frac{1}{\ell!}\left(w_{1}^{a_1-1}w_{2}^{a_2-1}\ldots w_{\ell}^{a_\ell-1}+w_{2}^{a_1-1}w_{1}^{a_2-1}\ldots w_{\ell}^{a_\ell-1}+\ldots\right)\,.
\end{align}
In this way, $\mathfrak{z}_+$ is completely symmetric in all of its arguments, while $\mathfrak{z}_-$ is separately symmetric in the two sets of indices.

The integrals $\mathfrak{z}_{+}(a_1,\ldots,a_k)$ and $\mathfrak{z}_{-}(a_1,\ldots,a_\ell|b_1\ldots,b_{k-\ell})$ are convergent if all the arguments are $>1$ (for numerical plots see Figures~\ref{Fig:NumericalSingleZeta}, \ref{Fig:NumericalDoubleZeta} and \ref{Fig:NumericalTrippleZeta} for depths $r=1,2,3$ respectively), but else does not converge at the origin. In this case, we understand  the integrals to be regularised by a cutoff $\epsilon\ll 1$ at the origin (see (\ref{RegularisationEpsilon}))
\begin{align}
\int_0^{\infty}d^n[\underline{t}]\longrightarrow \int_\epsilon^{\infty}d^n[\underline{t}]
\end{align}
Rewriting $\mathcal{Z}_{\emptyset}^{(2s),r}$ in (\ref{DepthSubDivision}) in terms of the $\mathfrak{z}_{+}(a_1,\ldots,a_k)$ and $\mathfrak{z}_{-}(a_1,\ldots,a_\ell|b_1\ldots,b_{k-\ell})$ yields a finite limit $\epsilon\to 0$ (\emph{i.e.} logarithmic singularities stemming from the regularisation, cancel among the various contributions (\ref{GenFormZetaContributions})). This allows us to write $\mathcal{Z}_{\emptyset,k}^{(2s),r}$ as combinations of finitely many $\mathfrak{z}_\pm$ and Riemann zeta functions, compatible with the depth $r$ (\emph{i.e.} the number of $\zeta_{a}$ and the number of arguments in $\mathfrak{z}_\pm$ add up to $r$) which is very schematically represented as (see also (\ref{CartoonContrIntro}))
\begin{align}
\mathcal{Z}_{\emptyset,k}^{(2s),r}\sim \sum C(\rho) \underbrace{\zeta\ldots\zeta}_{(r-k)\text{ factors}} \times \underbrace{\mathfrak{z}_\pm(\ldots)}_{k\text{-arguments}}\,.\label{CartoonContr}
\end{align}
Concrete expressions at $r=1$ are given in (\ref{SingleDepthZetaFunction}), at $r=2$ in (\ref{IntRepS0Term}), (\ref{InterS1Depth1}), (\ref{TplusContDepth2}) and (\ref{TminusContDepth2}), and at $r=3$ in (\ref{ConstTermDepth3}), (\ref{Depth3InterS2Contri}), (\ref{Depth3TinterContri}) and (\ref{ExplDepth3S1}). These examples in particular highlight that the schematic coefficients $C$ in (\ref{CartoonContr}) are polynomials in $g$ and depend in a highly non-trivial fashion on $r$ and $s$. Since the $\mathfrak{z}_\pm$-functions satisfy recursive relations, which allow to shift their arguments (see \emph{e.g.} (\ref{SingleZetReg}) and (\ref{Zeta2Identity})), it is possible to provide more compact expressions of (\ref{CartoonContr}), such as (\ref{ZpartZeta}) at $r=1$ and (\ref{FullDepth2Contr}) at $r=2$: these rewritings replace finite combinations of $\mathfrak{z}_\pm$ with differential operators acting on a single such function. While such rewritings are generally not unique, they structurally remind of algebraic relations in the context of multi-zeta values (see \emph{e.g.} \cite{Brown:2011ik,Brown:2011uzk,Broadhurst:1996kc,Zagier1994}). This link is further reinforced by observing that the individual contributions in (\ref{CartoonContr}) also respect a weight-structure, \emph{i.e.} terms proportional to an explicit power of $g$ have the same weight, which is defined as the sum of the arguments of the zeta functions and the $\mathfrak{z}_\pm$-functions. It would be interesting in the future to further investigate this potential relation. 

In the context of the perturbative partition function $\mathcal{Z}_\emptyset$, the rewriting in terms of Gauss-deformed zeta functions allows to spot certain recursive  patterns. For example, (\ref{ProductDepth2}) at $r=2$ and (\ref{Depth3TinterContri}) at $r=3$ suggest relations of the form
\begin{align}
\mathcal{Z}_{\emptyset,r-1}^{(2s),r}\sim\sum_{n=0}^{s-2}\frac{\zeta_{2n+1}}{n+1}\,\mathcal{Z}_{\emptyset,r-1}^{(2s-2n-2),r-1} \,,\label{IterativeRelations}
\end{align}
up to suitable combinatorical factors (and potential additive terms depending only on Riemann zeta functions that are required to remove spurious singularities from $\mathfrak{z}$-functions with arguments 1). The relation (\ref{Depth3InterS2Contri}) at depth 3 further suggests that these relations can be iterated, such that at any given depth, the only 'new' contributions stem from $\mathcal{Z}_{\emptyset,r}^{(2s),r}$, which are made up from Gauss-deformed zeta functions with the maximal number of arguments. 

\section{Leading Mass Contributions to the Perturbative Partition Function}\label{Sect:MassCorrection}
While the discussions of Section~\ref{Sect:MassOrder} were focused on different well-defined contributions $\mathcal{Z}_{\emptyset}^{(2s),r}$ at generic mass-order $\mathcal{O}(\madj^{2s})$ (but for fixed depth $r$), in this Section, for completeness, we shall discuss the full order $\mathcal{O}(\madj^{2s})$ of the perturbative partition function $\mathcal{Z}_{\emptyset}^{(2s)}$ for $s=1$ and $s=2$, thereby combining the different elements which we have worked out in the previous Section.
\subsection{Order $\mathcal{O}(\madj^2)$}\label{Sect:LeadingMassDeformation}
We first consider the case $s=1$, which captures the leading mass deformation of the perturbative partition function at order $\mathcal{O}(\madj^2)$. Following (\ref{IntroZ2sr}), at this order, the perturbative partition function consists only of the depth 1 contribution discussed in Section~\ref{Sect:Depth1Contribution}.
\begin{align}
\mathcal{Z}_{\emptyset}^{(2)}=\mathcal{Z}_{\emptyset}^{(2),1}&=-\sum_{k=1}^\infty(-1)^k\frac{\Gamma(k+3/2)}{\rho^{k+3/2}}\,2^{k+1}\frac{\zeta_{2k+1}}{k+1}\,\big(^{2k+2}_{\hspace{0.3cm}2}\big)\,.\label{Depth1SeriesS1}
\end{align}
Specifically, (\ref{ZetaResultZ2}) gives the explicit form of the regularised partition function (with the combination $(\ztu{1}-\zeta_1)$ explained in (\ref{AbusiveDefZeta1})), which can be written more compactly as
\begin{align}
\mathcal{Z}_{\emptyset}^{(2),1}=-\frac{\sqrt{\pi}}{\rho^{3/2}}\left[(\dop+1)(\dop+1)\ztu{1}-\zeta_{1}\right]\,.\label{ZpartZeta}
\end{align}
where $\dop$ is defined in (\ref{ShiftZetaValue}) and $\ztu{a}$ in (\ref{SingleZetReg}).\footnote{More specifically, $\ztu{1}$ (and $\zeta_1$) are understood as regularised integrals (\ref{RegularisationEpsilon}), such that the limit $\epsilon\to 0$ in (\ref{ZpartZeta}) is well-defined} 

\begin{figure}
\vspace{-0.25cm}
\begin{center}
\includegraphics[width = 7.5cm]{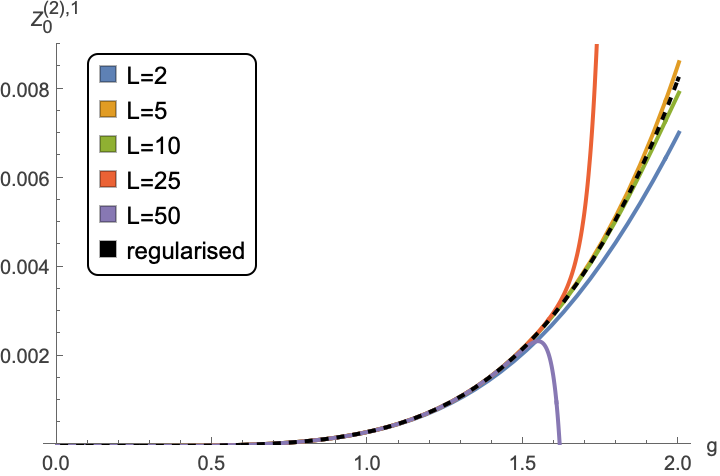}\hspace{1cm}\includegraphics[width = 7.5cm]{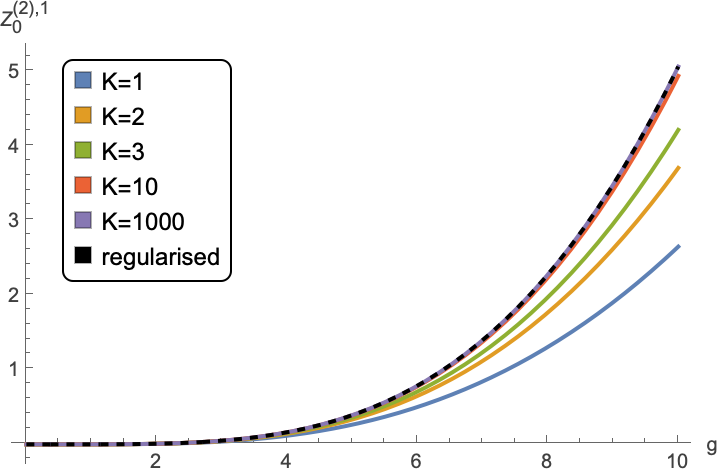}
\end{center}
\vspace{-0.5cm}
\caption{Numerical plot of $\mathcal{Z}_\emptyset^{(2),1}=\mathcal{Z}_\emptyset^{(2)}$ as a function of $g$. Left panel: truncated asymptotic series (\ref{SeriesZFiniteL}) for different values of $L$; Right panel: truncated Borel resummation (\ref{BorelResumFunZ01Trun}) for different values of $K$. In both plots, the black dashed line represents (\ref{Z1FormInteg}) (as a numerical evaluation of (\ref{ZetaResultZ2})), \emph{i.e.} a regularisation of the asymptotic series (\ref{SeriesZFiniteL}) using Gauss-deformed zeta functions.}
\label{Fig:NumericalZ1}
\vspace{-0.25cm}
\end{figure}

From (\ref{BorelResumSingleDepth}) we have the Borel resummed form of (\ref{Depth1SeriesS1})
\begin{align}
&\mathcal{Z}_{\emptyset}^{(2),1}=\mathcal{Z}_{\emptyset}^{(2)}=-\frac{g^3}{16\pi^2\sqrt{2\pi}}\,\sum_{\ell=1}^\infty\frac{1}{\ell}\,\left(g\,\tfrac{d}{dg}+1 \right)^2\,\ser\left(\tfrac{g^2}{\ell^2}\right)\,.\label{DefG02}
\end{align}
Using the expresssion (\ref{DerivativesErrorFunction}) for the derivative of the error-function, we may compute
\begin{align}
\frac{1}{\ell} \left(g\frac{d}{dg}+1\right)\mathcal{L}^0[\mathcal{B}[\ser]](g^2/\ell^2)&=\frac{16\pi^{7/2}\,\ell^2}{g^3}\,e^{\frac{4\pi^2\ell^2}{g^2}}\,\left(\text{Erf}\left(\tfrac{2\pi\ell}{g}\right)-1\right)+\frac{8\pi^2\ell}{g^2}-\frac{1}{\ell}\,,\nonumber\\
\frac{1}{\ell} \left(g\frac{d}{dg}+1\right)^2\mathcal{L}^0[\mathcal{B}[\ser]](g^2/\ell^2)&=-\frac{32\pi^{7/2}\ell^2}{g^3}\,e^{\frac{4\pi^2\ell^2}{g^2}}\,\left(\frac{4\pi^2\ell^2}{g^2}+1\right)\,\left(\text{Erf}\left(\tfrac{2\pi\ell}{g}\right)-1\right)\nonumber\\
&\hspace{0.75cm}-\frac{64\pi^4\ell^3}{g^4}-\frac{8\pi^2\ell}{g^2}-\frac{1}{\ell}\,,
\end{align}
which allows to write the Borel resummation of $\mathcal{Z}_{\emptyset}^{(2)}$ in the following form
\begin{align}
\mathcal{Z}_{\emptyset}^{(2)}\sim\sum_{\ell=1}^\infty\frac{1}{16\pi^2\sqrt{2\pi}} \bigg(&32\pi^{7/2}\ell^2\,e^{\frac{4\pi^2\ell^2}{g^2}}\,\left(\frac{4\pi^2\ell^2}{g^2}+1\right)\,\left(\text{Erf}\left(\tfrac{2\pi\ell}{g}\right)-1\right)+\frac{64\pi^4\ell^3}{g}+8\pi^2\ell g+\frac{g^3}{\ell}\bigg)\,.\label{BorelResumFunZ2}
\end{align}
This correctly reproduces the series (\ref{Depth1SeriesS1}) for $g\to 0$ up to exponentially suppressed terms. In order to verify this, we use the expansion of the error function in (\ref{ErfExpansion})
\begin{align}
\mathcal{Z}_{\emptyset}^{(2)}&\sim-\sum_{\ell=1}^\infty\frac{1}{16\pi^2\sqrt{2\pi}} \bigg(32\pi^{7/2}\ell^2\,\left(\frac{4\pi^2\ell^2}{g^2}+1\right)\,\mathcal{E}\left(\tfrac{2\pi\ell}{g}\right)-\frac{64\pi^4\ell^3}{g}-8\pi^2\ell g-\frac{g^3}{\ell}\bigg)\nonumber\\
&=-\sum_{\ell=1}^\infty\frac{1}{16\pi^2\sqrt{2\pi}} \bigg(32\pi^{3}\ell^2\,\left(\frac{4\pi^2\ell^2}{g^2}+1\right)\,\left(\sum_{k=1}^\infty (-1)^{k-1}\,\frac{(2k-3)!!}{2^{k-1}}\,\left(\frac{g}{2\pi\ell}\right)^{2k-1}\right)\nonumber\\
&\hspace{3.75cm}-\frac{64\pi^4\ell^3}{g}-8\pi^2\ell g-\frac{g^3}{\ell}\bigg)\,.
\end{align}
Using the identity $(2k-3)!!=\frac{(2k-2)!}{2^{k-1}(k-1)!}$ and (\ref{DefZetaFunction}) we find
{\allowdisplaybreaks
\begin{align}
\mathcal{Z}_{\emptyset}^{(2)}&\sim\frac{1}{\sqrt{2\pi}}\sum_{\ell=1}^\infty\bigg[64\pi^4\ell^5\sum_{k=4}^\infty(-1)^k\,\frac{(2k-2)!}{(k-1)!}\,\frac{g^{2k-3}}{(16\pi^2\ell^2)^k}+16\pi^2\ell^3\sum_{k=3}^\infty(-1)^k\,\frac{(2k-2)!}{(k-1)!}\,\frac{g^{2k-1}}{(16\pi^2\ell^2)}\bigg]\nonumber\\
&=-\frac{1}{16\pi^2\sqrt{2\pi}}\sum_{k=1}^\infty (-1)^k\,\zeta_{2k+1} \,(2k+1)^2 \, \frac{(2k)!}{k!}\,\frac{g^{2k+3}}{(16\pi^2)^k}\,.
\label{CheckResummed}
\end{align}}
This in fact reproduces the initial series in (\ref{Depth1SeriesS1}), such that indeed (\ref{BorelResumFunZ2}) represents a resummation of $\mathcal{Z}_{\emptyset}^{(2)}$ (up to exponentially suppressed terms). In order to visualise this result relative to (\ref{Depth1SeriesS1}), we defined the truncated series
\begin{align}
\mathcal{Z}_{\emptyset}^{(2),1}(L):=\sum_{k=1}^L(-1)^{k+1}(2k+1)2^{k+1}\zeta_{2k+1}\,\frac{\Gamma(k+3/2)}{\rho^{k+3/2}}\,.\label{SeriesZFiniteL}
\end{align}
This expression is plotted as a function of $g$ for different values of $L$ in the left panel of Figure~\ref{Fig:NumericalZ1} and compared to the regularised result (\ref{ZetaResultZ2}). As is visible from this plot, for finite values of $g$, increasing $L$ initially improves the convergence (towards the resummed result (\ref{BorelResumFunZ2})). However, for too large values of $L$, the divergence of the series becomes apparent: in fact, the point at which the expression $\mathcal{Z}_{\emptyset}^{(2),L}$ starts strongly deviating is shifted towards smaller values of $g$. Before continuing, we remark on the convergence of the series in $\ell$ in (\ref{BorelResumFunZ2}). Indeed, the last three terms in the bracket of these expression taken alone do not constitute convergent series. However, they are present to compensate similar divergent contributions stemming from the error function to render the entire series convergent. Indeed, this is visible in (\ref{CheckResummed}), where the summation over $\ell$ is a convergent in both terms. We have illustrated this  fact in the right panel of Figure~\ref{Fig:NumericalZ1}, where, in fact, we have shown a truncation of the series over $\ell$ up to 1000 terms:
\begin{align}
\mathcal{Z}_{\emptyset}^{(2),1}(K):=\sum_{\ell=1}^K\frac{1}{16\pi^2\sqrt{2\pi}} \bigg(&32\pi^{7/2}\ell^2\,e^{\frac{4\pi^2\ell^2}{g^2}}\,\left(\frac{4\pi^2\ell^2}{g^2}+1\right)\,\left(\text{Erf}\left(\tfrac{2\pi\ell}{g}\right)-1\right)+\frac{64\pi^4\ell^3}{g}+8\pi^2\ell g+\frac{g^3}{\ell}\bigg)\,.\label{BorelResumFunZ01Trun}
\end{align}
As can be observed in the right panel of Figure~\ref{Fig:NumericalZ1}, increasing $K$ improves the agreement with the regularised result (\ref{Z1FormInteg}).

\subsection{Order $\mathcal{O}(\madj^4)$}\label{Sect:OrderEps4}
We next consider the (full) contribution $\mathcal{Z}_\emptyset^{(4)}$ at order $\mathcal{O}(\madj^4)$, which has two parts (see 

\begin{wrapfigure}[13]{r}{0.43\textwidth}
\vspace{-0.5cm}
\begin{center}
\includegraphics[width = 7.1cm]{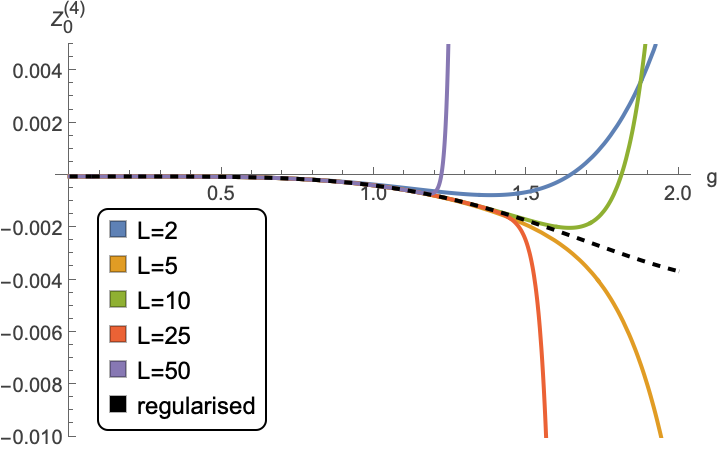}
\end{center}
\vspace{-0.5cm}
\caption{Numerical plot of $\mathcal{Z}_\emptyset^{(4)}$: the coloured lines show $\mathcal{Z}_\emptyset^{(4),1}(L)+\mathcal{Z}_{\emptyset}^{(4),2}(L)$ (see (\ref{Truncates21}) and (\ref{Truncates22})) for different values of $L$. The dashed black line shows the regularised contribution.}
\label{Fig:FullZ04}
\vspace{-0.5cm}
\end{wrapfigure}
\begin{figure}
\begin{center}
\includegraphics[width = 7.5cm]{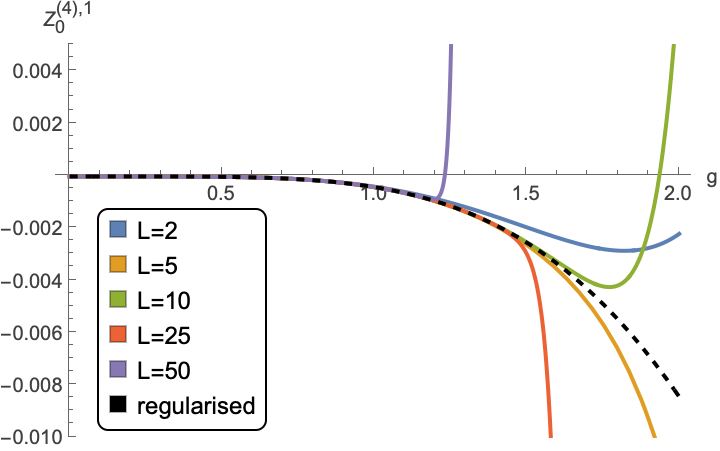}\hspace{1cm}\includegraphics[width = 7.5cm]{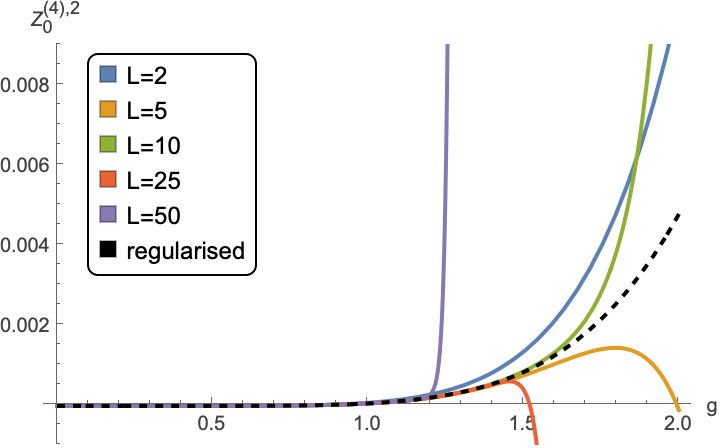}
\end{center}
\caption{Numerical plot of the two contributions $\mathcal{Z}_\emptyset^{(4),1}$ (left) and $\mathcal{Z}_\emptyset^{(4),2}$ (right) in (\ref{SplitZ04}) as functions of $g$. Left panel: the coloured lines represent the truncated asymptotic series (\ref{Truncates21}) for different values of $L$, while the dashed black lines provides the regularised result (\ref{SeparateDepths4}). Right panel: the coloured lines represent the truncated asymptotic series (\ref{Truncates22}) for different values of $L$, while the dashed black lines provides the regularised result (\ref{SeparateDepths4}).} 
\label{Fig:NumericalZ2Compar}
\vspace{-0.5cm}
\end{figure}
\begin{figure}
\begin{center}
\includegraphics[width = 7.5cm]{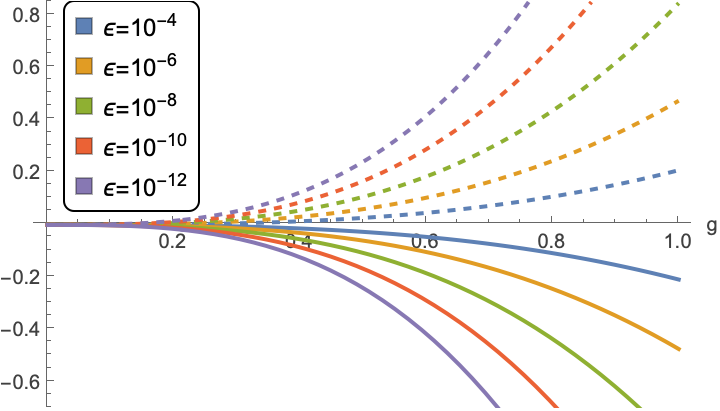}\hspace{1cm}\includegraphics[width = 7.5cm]{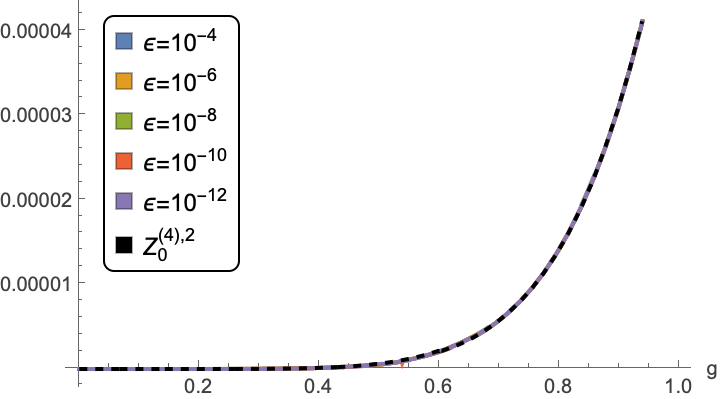}
\end{center}
\caption{Left panel: Numerical plot of $\mathcal{Z}_{\emptyset,0}^{(4),2}+\mathcal{Z}_{\emptyset,1}^{(4),2}$ (solid lines) and $\mathcal{Z}_{\emptyset,2}^{(4),2}$ where divergent contributions are regularised by a cutoff $\epsilon\ll 1$. Right panel: combined contribution $\mathcal{Z}_{\emptyset,0}^{(4),2}+\mathcal{Z}_{\emptyset,1}^{(4),2}+\mathcal{Z}_{\emptyset,2}^{(4),2}$ for different values of the regulator. The dashed black line corresponds to the expression where all three contributions have been written as a single (convergent) integral without regulator at the origin (which is the same as in the right panel of Figure~\ref{Fig:NumericalZ2Compar}).} 
\label{Fig:NumericalZ2Cutoff}
\vspace{-0.25cm}
\end{figure}


\noindent
the definition (\ref{IntroZ2sr})), namely depth 1 and depth 2:
\begin{align}
\mathcal{Z}_\emptyset^{(4)}=\mathcal{Z}_\emptyset^{(4),1}+\mathcal{Z}_\emptyset^{(4),2}\,.\label{SplitZ04}
\end{align}
A graphic representation of the full function $\mathcal{Z}_\emptyset^{(4)}$ is shown in Figure~\ref{Fig:FullZ04}. More specifically, the coloured lines in this Figure show the truncated series $\mathcal{Z}_\emptyset^{(4),1}(L)+\mathcal{Z}_\emptyset^{(4),2}(L)$ for different values of $L$, where we have introduced truncated versions of the depth 1 contribution (\ref{Depth1Series}) 
\begin{align}
\mathcal{Z}_\emptyset^{(4),1}(L)=\sum_{k=1}^L&\frac{(-1)^k\,2^{k-1}\,(2k+3)\,(2k+2)\,(2k+1)}{3\rho^{k+3/2}}\nonumber\\
&\times\Gamma(k+3/2)\,\zeta_{2k+3}\,.\label{Truncates21}
\end{align}
and of the depth 2 contribution (\ref{Depth2Series})
\begin{align}
\mathcal{Z}_{\emptyset}^{(4),2}(L)&=\sum_{k=2}^L (-1)^k \frac{\Gamma(k+3/2)}{\rho^{k+3/2}}\,2^{k+1}\sum_{\ell=1}^{k-1}\sum_{n=\ell}^{\ell}\frac{\zeta_{2n+1}}{n+1}\,\frac{\zeta_{2k-2n-1}}{k-n+1}\,\big(^{\hspace{0.15cm}2(n+1)}_{2(n-\ell+1)}\big)\,\big(^{2(k-n+1)}_{\hspace{0.05cm}2(\ell-n+1)}\big)\,.\label{Truncates22}
\end{align}
For simplicity (given the complicated nature of the Borel resummed result (\ref{Depth2SerRewrite})), we shall here only discuss the regularisation of $\mathcal{Z}_\emptyset^{(4)}$. Indeed, the dashed line in Figure~\ref{Fig:FullZ04} shows the regularised expression $\mathcal{Z}_\emptyset^{(4)}$, which is comprised of 
\begin{align}
&\mathcal{Z}_\emptyset^{(4),1}=\frac{\sqrt{\pi}}{2\rho^{3/2}}\left[\tfrac{1}{3}\,(\dop+1)(\dop+3)\ztu{3}-\tfrac{1}{2}\,\zeta_{3}\right]\,,&&\text{and}&&\mathcal{Z}_\emptyset^{(4),2}=\mathcal{Z}_{\emptyset,0}^{(4),2}+\mathcal{Z}_{\emptyset,1}^{(4),2}+\mathcal{Z}_{\emptyset,2}^{(4),2}\,,\label{SeparateDepths4}
\end{align}
which are separately plotted and compared to the truncated series (\ref{Truncates21}) and (\ref{Truncates22}) in Figure~\ref{Fig:NumericalZ2Compar}.
Specifically, the first expression in (\ref{SeparateDepths4}) is (\ref{ZpartZeta}) for the contribution of depth $1$, while the second part uses the notation (\ref{DepthSubDivision}) to write the contribution at depth 2 with
\begin{align}
\mathcal{Z}_{\emptyset,0}^{(4),2}+\mathcal{Z}_{\emptyset,1}^{(4),2}&=-\frac{\sqrt{\pi}\, \zeta_{1}}{\rho^{3/2}}\,\left(2\,(\dop+1)(\dop+1)\,\ztu{1}-\zeta_{1}\right)\,,\nonumber\\
\mathcal{Z}_{\emptyset,2}^{(4),2}&=\frac{\sqrt{\pi}}{\rho^{5/2}}\left[\rho\,(\dop+1)\left(\dop+1\right)\ztua{1}{1}+(\dop+1)(\dop+3)\,\ztub{2}{2}\right]\,.\label{Z42decomp}
\end{align}
Here, $\mathcal{Z}_{\emptyset,2}^{(4),2}$ contains the contributions of $\mathfrak{z}$-functions with 2 arguments (\emph{i.e.} defined in (\ref{DefZPlus}) and (\ref{DefZMinus})), while $\mathcal{Z}_{\emptyset,0}^{(4),2}+\mathcal{Z}_{\emptyset,1}^{(4),2}$ contains contributions of $\mathfrak{z}$-functions of a single argument (defined in (\ref{SingleZetReg})) and Riemann zeta function (defined in (\ref{DefZetaFunction})). Figure~\ref{Fig:NumericalZ2Compar} demonstrates once more the asymptotic nature of the series (\ref{Depth1Series}) and (\ref{Depth2Series}). Furthermore, this Figure also shows that both $\mathcal{Z}_{\emptyset}^{(4),1}$ and notably $\mathcal{Z}_{\emptyset}^{(4),2}$ are finite functions of $g$. This is non-trivial in the case of the latter, since both terms in (\ref{Z42decomp}) contain formally divergent elements, such as the divergent Riemann zeta function $\zeta_{1}$ and the Gauss-deformed versions $\ztu{1}$ and $\ztua{1}{1}$: the combination of these terms is such that possible divergences cancel. This is shown numerically in Figure~\ref{Fig:NumericalZ2Cutoff}. In all terms, the singularity stems from a divergence of integrals at the origin (see the discussion at the end of Section~\ref{Sect:Depth1Reg}). Regularising this contribution with a cutoff $\epsilon\ll 1$ (see (\ref{EpsilonRegularisation})) leads to finite contributions, which are individually plotted in the left panel of Figure~\ref{Fig:NumericalZ2Cutoff} as a function of the cutoff $\epsilon$: both $\mathcal{Z}_{\emptyset,0}^{(4),2}+\mathcal{Z}_{\emptyset,1}^{(4),2}$ and $\mathcal{Z}_{\emptyset,2}^{(4),2}$ grow (logarithmically) as $\epsilon\to 0$, thus indicating a divergence of each individual term. However, the combination of these contributions is finite for any value of the cutoff and indeed yields $\mathcal{Z}_0^{(4),2}$ as is shown in the right panel of Figure~\ref{Fig:NumericalZ2Cutoff}.

\section{Conclusions}\label{Sect:Conclusions}
In this paper we have studied the perturbative partition function of $\mathcal{N}=2^*$ SYM theory on $\mathbb{S}^4$. Using localisation techniques, this function is written essentially as a Gaussian integral over the 1-loop determinant $\Delta_{\text{1-loop}}$ in (\ref{OneLoopDet}). Expanding this contribution in the adjoint mass parameter $\madj$, we can compute the term at order $\mathcal{O}(\madj^{2s})$ (which we denote $\mathcal{Z}_\emptyset^{(2s)}$, see (\ref{InitialMassExpansion})) using the Laplace method. To further analyse the result, we have found it useful to further decompose a given $\mathcal{Z}_\emptyset^{(2s)}$ into contributions of different \emph{depth} (see (\ref{IntroZ2sr})). Indeed, an expansion of $\Delta_{\text{1-loop}}$ naturally organises the contributions to the saddle-point approximation into terms containing products of precisely $r$ Riemann zeta functions (where $r\in\{0,1,\ldots,s\}$). Using nomenclature from the study of multi-zeta values \cite{Zagier1994,Brown:2011ik,Brown:2011uzk,Broedel_2013}, we therefore call the corresponding terms $\mathcal{Z}_\emptyset^{(2s),r}$ the contributions at \emph{depth} $r$. However, after integration, we find $\mathcal{Z}_\emptyset^{(2s),r}$ in the form of an asymptotic series in the coupling constant $g$, \emph{i.e.} a power series with vanishing radius of convergence. In order to still make sense of this object, we have employed two complementary strategies
\begin{itemize}
\item[\emph{(i)}] {\bf Borel resummation:} the series encountered in $\mathcal{Z}_\emptyset^{(2s),r}$ are Gevrey-1. Therefore, using techniques briefly reviewed in Appendix~\ref{App:ResurgenceReview}, we can Borel resum them into infinite (converging) sums over error functions (acted upon by differential operators in $g$). The examples at depth $r=1,2$ can be found in (\ref{BorelResumSingleDepth}) and (\ref{Depth2SerRewrite}). The appearance of error functions in a similar context has previously been noted in \cite{Aniceto:2026pcz}.

\item[\emph{(ii)}] {\bf regularisation:} replacing the Riemann zeta functions in $\mathcal{Z}_\emptyset^{(2s),r}$ by Mellin integrals with a cutoff parameter turns the asymptotic series into a convergent one. In the limit where the cutoff reaches infinity, we obtain a regularised version of $\mathcal{Z}_\emptyset^{(2s),r}$. Using this strategy for $s=1=r$, the partition function has previously been written in a (convergent) integral representation. Here we argue that systematically, for generic mass order $s$ and depth $r$, we can write $\mathcal{Z}_\emptyset^{(2s),r}$ in terms of finitely many new functions (see (\ref{SingleZetReg}) for the initial definition as well as (\ref{GenZetaGenDepthPlus}) and (\ref{GenZetaGenDepthMinus}) for generalisations to arbitrary depth). Due to their similarity to the Mellin representation of the Riemann zeta function, with an additional insertion of a Gaussian kernel, we call them Gauss-deformed zeta functions. We have given explicit expressions up to $r=3$, which we have checked up to $s=30$ for $r=1$, $s=20$ for $r=2$ and $s=12$ for $r=3$. We have also established a relation among the depth 1 $\mathfrak{z}$-functions and converging sums over error functions (see Figure~\ref{Fig:RelationAsymSeries} and Section~\ref{Sect:Correspondence}), thus rendering the connection to approach \emph{(i)} more tangible
\end{itemize}
The approaches are complementary and present different aspects of the mass deformations of the partition function. The rewriting in terms of Gauss-deformed zeta functions renders the perturbative part of the partition function more compact, since for given $s$, only a finite number of functions appear. Even among these functions, further non-trivial relations exist (see \emph{e.g.} (\ref{ShiftZetaValue}) or (\ref{Zeta2Identity})), which afford even more compact expressions in terms of a restricted set of Gauss-deformed zeta functions (acted upon by differential operators in the coupling $g$). Similar relations exist also at higher depth (\emph{i.e.} for the functions (\ref{GenZetaGenDepthPlus}) and (\ref{GenZetaGenDepthMinus})), which can systematically be used to simplify the higher depth contributions to $\mathcal{Z}_\emptyset^{(2s),r}$. However, any such rewriting is not unique and it is difficult to determine the 'simplest' form without any abstract guiding principle. This situation reminds of similar relations among different multi-zeta values (MZV): in this case a Hopf-algebra structure exists~\cite{Brown:2011ik,Brown:2011uzk}, which can guide the re-organisation of expressions involving multiple MZV, \emph{e.g.} in the calculation of scattering amplitudes in string- and gauge theory \cite{Terasoma_2002,Oprisa:2005wu,Stieberger:2009rr,Mafra:2011nv,Mafra:2011nw,Schlotterer:2012ny,Broedel:2013tta,Drummond:2013vz,Stieberger:2013wea,Stieberger:2016xhs,Brown:2018omk,Brown:2018omk,Brown:2009ta,Brown:2008um,Brown:2009qja,Duhr:2015rjo,Todorov:2014tya} (see also \cite{Kervyn:2026nal} for a recent application of resurgence techniques in this context). Based on this analogy, we speculate if a similar (algebraic) structuring principle exists in the case of the sphere partition function.

The compactness achieved for the expressions of the perturbative part of the partition function allows to highlight non-trivial recursive relations between contributions at various depths, as for example (\ref{ProductDepth2}) at depth 2 and (\ref{Depth3TinterContri}) and (\ref{Depth3InterS2Contri}) at depth 3. These patterns lead us to conjecture similar relations at all depths (see \emph{e.g.} (\ref{IterativeRelations})) and we expect that the truly 'new' information at each depth can be written entirely in terms of Gauss-deformed zeta functions with the maximal number of arguments allowed at any given order $s$ in the mass-expansion. This observation in fact, provides a more direct link between the depth (which so far is only a mathematical organisation principle for the partition function) and $s$ (which a priori carries physical significance as the order of the mass expansion). Once more, this situation reminds of the role of motives and multi-zeta values played in the study of scattering amplitudes~\cite{Broedel_2013,Schlotterer:2012ny}. We leave the study of a possible analogous structure underlying the sphere partition function for further studies.

Finally, in this paper we have exclusively discussed the perturbative part of the partition function. Although much more involved, the instanton corrections can be discussed in a very similar fashion. It will be interesting to see if these contributions can also be captured by the newly proposed deformed zeta functions (maybe with additional deformations of the Gauss-kernel). Similarly, the discussion presented here could be extended to the gauge theory on $\mathbb{S}^4\times \mathbb{S}^1$ and $\mathbb{S}^4\times T^2$, which would provide higher-dimensional generalisations of our approach. We leave these ideas for future work.


\section*{Acknowledgements}
We are very grateful to  Baptiste Filoche for numerous discussions and exchanges on related topics. TA and SH would like to thank the Institut de Mathématiques de Bourgogne (IMB) for kind hospitality while part of this work was being completed.
The work of TK is supported by EIPHI Graduate School (No.~ANR-17-EURE-0002) and the Bourgogne-Franche-Comté region.

\appendix

\section{Useful Special Functions}\label{App:Functions}
In this Appendix, we compile the definitions (and certain properties) of a number of special functions, which shall be useful in the main text of the text. 

\paragraph{Barnes $\mathsf{G}$-function, gamma function, and Riemann zeta function}
The 1-loop determinant $\Delta_{\text{1-loop}}$ in (\ref{OneLoopDet}) is expressed in terms of the Barnes $\mathsf{G}$-function \cite{Barnes}
\begin{align}
\mathsf{G}(1+z)=(2\pi)^{z/2}\,e^{-\frac{1}{2}((1+\gamma)z^2+z)}\prod_{n=1}^\infty\left(1+\frac{z}{n}\right)^n\,e^{-z+\frac{z^2}{2n}}\,,\label{DefBarnesDoubleGamma}
\end{align}
which can be combined into the function \cite{Pestun:2007rz}
\begin{align}
H(z)=\mathsf{G}(1+z)\mathsf{G}(1-z)=e^{-(1+\gamma)z^2}\,\prod_{n=1}^\infty\,\left(1-\frac{z^2}{n^2}\right)^n\,\prod_{n=1}^\infty e^{\frac{z^2}{n}}\,.
\end{align}
The $\mathsf{G}$-function affords the following Taylor series expansion around the origin (see, for example,~\cite{Voros})
\begin{align}
&\log \mathsf{G}(1+z)=(\log\,2\pi -1)\frac{z}{2}-(1+\gamma)\,\frac{z^2}{2}+\sum_{n=3}^\infty \frac{(-1)^{n-1}}{n}\,\zeta_{n-1}\,z^n\,,&&\text{for} &&|z|<1\,,
\end{align}
The expansion of $\Delta_{\text{1-loop}}$ in (\ref{DeltaFormExpansion}) gives rise to coefficients which can be related to the digamma-function as
\begin{align}
\psi(z)=\frac{\Gamma'(z)}{\Gamma(z)}\,,\label{Digamma}
\end{align}
with $\Gamma$ the gamma function (and $\Gamma'$ its derivative). For $|z|<1$ it satisfies the following series expansion
\begin{align}
&\psi(1+z)=-\gamma-\sum_{k=1}^\infty(-1)^k\,\zeta_{k+1}\,z^k\,,&&\text{for} &&|z|<1\,.\label{SeriesExpansionDigamma}
\end{align} 
More concretely, the expansion coefficients of $\Delta_{\text{1-loop}}$ in (\ref{FormckCoef}) also depend on the \emph{Riemann zeta function}, which is defined as
\begin{align}
&\zeta_s=\sum_{\ell=1}^\infty\,\ell^{-s}\,,&&\text{for} &&\text{Re}(s)>1\,.\label{DefZetaFunction}
\end{align}
The Mellin representation of the zeta function is given by
\begin{align}
\zeta_s=\frac{1}{\Gamma(s)}\int_0^\infty\frac{x^{s-1}}{e^x-1}\,dx\,,\label{MellinRepresentationZeta}
\end{align}
which can equivalently be written as
\begin{align}
\zeta_{2s+1}=\frac{2^{2s}}{\Gamma(2s+2)}\,\int_0^\infty dw\,\frac{w^{2s+1}}{\sinh^2 w}\,.\label{FormZetaIntegSinh}
\end{align}

We also define the \emph{Pochhammer symbol} as the finite propduct
\begin{align}
(x)_n=\prod_{k=0}^{n-1}(x+k)=\frac{\Gamma(x+n)}{\Gamma(x)}\,.\label{DefPochhammer}
\end{align}

\paragraph{Error function}
Another function that is important for computations in the main text is the \emph{error function}, which is defined as 
\begin{align}
\text{Erf}(z)=\frac{2}{\sqrt{\pi}}\,\int_0^ze^{-t^2}\,dt\,,&&\forall z\in\mathbb{C}\,,\label{DefErfFunction}
\end{align}
Its derivatives can be written in the form 
\begin{align}
&\frac{d\text{Erf}}{dz}(z)=\frac{2}{\sqrt{\pi}}\,e^{-z^2}\,,&&\text{and}&&\frac{d^n \text{Erf}}{dz^n}(z)=(-1)^{n-1}\,\frac{2}{\sqrt{\pi}}\,H_{n-1}(z)\,e^{-z^2}\hspace{1cm}\forall n\geq 2\,,\label{DerivativesErrorFunction}
\end{align}
with $H_n$ the Hermite polynomial. Around $z\sim 0$, the error-function affords the following series expansion
\begin{align}
\text{Erf}(z)=-\frac{2}{\sqrt{\pi}}\sum_{k=1}^\infty\frac{(-1)^k z^{2k-1}}{(k-1)! (2k-1)}\,.\label{SeriesErrorFct}
\end{align}
while around $z^{-1}\sim 0$ we may write \cite{NIST}
\begin{align}
\text{Erf}(z)&=_{z\to\infty}1-\frac{e^{-z^2}}{\sqrt{\pi}}\underbrace{\left(z^{-1}-\frac{z^{-3}}{2}+\frac{3z^{-5}}{4}-\frac{15z^{-7}}{8}+\mathfrak{o}(z^{-7})\right)}_{=\mathcal{E}(z)}\,.\label{ErfExpansion}
\end{align}
Here $\mathcal{E}$ can formally be written as the following asymptotic series
\begin{align}
\mathcal{E}(z)=\sum_{n=1}^\infty (-1)^{n-1}\,\frac{(2n-3)!!}{2^{n-1}}\,z^{-2n+1}\,.\label{SerExpErf}
\end{align}
We also remark that $\text{Erf}$ is an odd function, \emph{i.e.} $\text{Erf}(-z)=-\text{Erf}(z)$.

\section{Resurgence}
\subsection{Review of Resurgence}\label{App:ResurgenceReview}
One of the approaches used in our work to give sense to asymptotic series draws heavily from ideas that have been developed within the framework of resurgence. For completeness, we give a brief, self-contained review of resurgence in this Appendix. For a more complete discussion, we refer the reader to the excellent reviews in the literature \cite{Marino:2012zq,Dorigoni:2014hea,Aniceto:2018bis}.

Consider the formal power series in $g\in\mathbb{R}$
\begin{align}\label{eq:res-formal}
&\widetilde\Phi(g)=\sum_{n=0}^{\infty}a_n g^{n}\,,&&\text{with} &&a_n\in\mathbb{C}\,.
\end{align}
A function $F(g)$ is approximated to successively
higher orders as $g\to0$ by $\widetilde \Phi$ if, after truncation at order $N$, the remainder is $\mathcal{O}(g^{N+1})$ for every
fixed $N$ : 

\begin{align}
    F(g) - \sum_{n=0}^{N} a_n g^{n} = \mathcal{O} (g^{N+1})
\end{align}
The series \eqref{eq:res-formal} is called \emph{Gevrey-1} if there exist
constants $C,K>0$ such that
\begin{equation}\label{eq:res-gevrey}
 |a_n|\leq C K^n n!,\qquad n\geq0.
\end{equation}
This is the characteristic factorial growth often found in perturbation theory in physical problems. The stronger statement that $F$ \emph{has a Gevrey-1 asymptotic
expansion} means that its truncation error obeys
\begin{equation}\label{eq:res-remainder}
 \left|F(g)-\sum_{n=0}^{N}a_n g^{n}\right|
 \leq C' (K')^N N!\,|g|^{N+1}
\end{equation}

To any formal series $\widetilde{\Phi}$ in (\ref{eq:res-formal}), we can assign another formal series where the factorial growth is removed by the \emph{Borel transform}

\begin{align}
\mathcal{B}:\,\mathbb{C}\llbracket g \rrbracket&\longrightarrow \mathbb{C}\llbracket \xi \rrbracket\nonumber\\
g^n&\longmapsto\tfrac{\xi^n}{n!}\,,\label{BorelTransform}
\end{align}
where $\mathbb{C}\llbracket z \rrbracket$ denotes the set of formal power series in $z$. More concretely, we have for the Borel transform of (\ref{eq:res-formal})
\begin{align}
    \mathcal{B}[\widetilde\Phi](\xi) := \sum_{n=0}^{+\infty} \frac{a_n}{n!}\,\xi^n\,,
\end{align}
which has a non-zero radius of convergence if $\widetilde \Phi $ is Gevrey-1. If the Borel transform has no singularities along a ray $\{r\,e^{i\theta}|r\in[0,\infty)\}$, of certain angle $\theta\in[0,2\pi)$, and has sub-exponential growth at infinity, the following object called the \emph{Borel-Laplace sum} (also called directional Laplace transformation) is a well-defined function for sufficiently small $g$:
\begin{align}\label{eq:res-laplace}
    \mathcal{L}^{\theta}[\mathcal{B}[\widetilde\Phi]] (g):= \frac{1}{g}\int_0^{e^{i\theta}\infty} e^{-\xi/g} \mathcal{B}[\widetilde\Phi](\xi) d\xi\,.
\end{align}
If well-defined, $\mathcal{L}^{\theta}[\mathcal{B}[\widetilde\Phi]] $ permits a Gevrey-1 asymptotic expansion that corresponds to $\widetilde \Phi$. If $\mathcal{B}[\widetilde\Phi]$ (or more precisely its analytic continuation) has singularities, the directional Laplace transformation depends on the angle $\theta$. Indeed, rays passing through the singular points (\emph{e.g.} poles, or logarithmic/power-law branch cuts) of $\mathcal{B}[\widetilde\Phi]$ cannot be used to define \ref{eq:res-laplace} and are called \emph{Stokes rays}. Given a stokes ray along a direction $\alpha$, the directional Laplace transforms with $\alpha_+ =\alpha +\epsilon$ and $\alpha_- =\alpha-\epsilon$ lead to different resummations. These two contributions, for $\epsilon \to 0$, differ by a Stokes discontinuity. Concretely, in the case of a simple pole (\emph{i.e.} $ \mathcal{B}[\widetilde \Phi](\xi) \sim \frac{-1}{2\pi i}\frac{\kappa}{\xi-A} $):
\begin{align} \label{eq:res-discontinuity}
(\mathcal{L}^{\alpha_+}- \mathcal{L}^{\alpha_-})\widetilde \Phi(g)= \kappa e^{-A/g}\,.
\end{align}
The discontinuity also has an explicit expression in the case of branch cuts \cite{Marino:2012zq}. 
The underlying interplay between perturbation theory and non-perturbative effects was investigated extensively in quantum mechanics, notably in the studies of the quartic harmonic oscillator \cite{Voros1983} and instanton-anti-instanton configurations \cite{BOGOMOLNY1980431,ZINNJUSTIN1981125}. Related ideas were subsequently developed in quantum field theory, including the semiclassical resurgence program reviewed by Dunne and \"Unsal \cite{DunneUnsalWKB2014,Dunne:2016nmc}, and in matrix models and topological strings, where non-perturbative saddles govern the large-order behavior of genus expansions \cite{Marino:2007te,Aniceto:2011nu}. 

\subsection{Borel Resummation of a Series}\label{App:BorelResumSer}
As a concrete example of an asymptotic series, which plays an important role in the main part of this paper, we consider the following series
\begin{align}
&\ser (z)= \sum_{k=1}^\infty \alpha_{k}\,z^{k}\,,&&\text{with} &&\alpha_{k}=\frac{(-1)^k(2k)!}{k!\, (16\pi^2)^k}\,.\label{SeriesInitial}
\end{align}
For $k\to\infty$ we have
\begin{align}
&\left|\frac{\alpha_{k+1}}{\alpha_{k}}\right|\sim_{k\to \infty} \frac{k}{4\pi^2}\,,&&\text{and} &&\alpha_{k}\sim_{k\to\infty} \frac{(-1)^k\,k!}{(4\pi^2)^k\,\sqrt{\pi k}}\,,
\end{align}
which indicates that $\ser$ has zero radius of convergence and is of Gevrey-1 type \cite{Dorigoni:2014hea}. The Borel transform (\ref{BorelTransform}) of $\ser$ than takes the following form
\begin{align}
\mathcal{B}[\ser](s)=\sum_{k=1}^\infty \frac{(-1)^k(2k)!}{(k!)^2}\left(\frac{s}{16\pi^2}\right)^k\,.\label{BorelTrafoSer}
\end{align}
Since $\frac{(2k)!}{(k!)^2}\sim_{k\to\infty}\frac{4^k}{\sqrt{\pi k}}$, the sum over $k$ in $\mathcal{B}[\ser](s)$ converges (absolutely) for $ |s| < 4\pi^2$, such that
\begin{align}
&\mathcal{B}[\ser](s)= \frac{1}{\sqrt{1+\tfrac{s}{4\pi^2}}} -1\,,&&\text{for}&&|s| < 4\pi^2\,.\label{BorelSum}
\end{align}
Here we have used
\begin{align}
&\sum_{k=0}^\infty\left(\begin{linesmall}{@{}c@{}} \\[-6pt]2k \\[3pt] k \end{linesmall}\right) z^k=\frac{1}{\sqrt{1-4z}}\,,&&\forall |z|<\tfrac{1}{4}\,.
\end{align}
The function $\mathcal{B}[\ser]$ is singular at $s_0=-4\pi^2$ but is finite for all $s$ with $\text{Re}(s)>0$. Therefore, in order to obtain a resummation of $\ser$, we calculate the (directional) Laplace transform (see Appendix~\ref{App:ResurgenceReview} for a review) 
\begin{align}
\mathcal{L}^0[\mathcal{B}[\ser]](z)=\frac{1}{z}\int_0^\infty e^{-\frac{s}{z}}\,\mathcal{B}[\ser](s)\,d s =\underbrace{\int_0^{\infty}\frac{e^{-s/z}\,d s}{z\sqrt{1+\frac{s}{4\pi^2}}}}_{=\mathcal{I}}-\underbrace{\int_0^\infty \frac{e^{-\frac{s}{z}}}{z}\,d s}_{=1}\,.\nonumber
\end{align}
In order to calculate the integral $\mathcal{I}$, we change variables $s=4\pi^2(t^2-1)$ (such that $ds=8\pi^2\,t\,dt$)
\begin{align}
\mathcal{I}=\frac{1}{z}\int_{1}^\infty e^{-\frac{4\pi^2}{z}(t^2-1)}=\frac{e^{\frac{4\pi^2}{z}}}{z}\,2\sqrt{\pi^2 z}\,\left(1-\text{Erf}\left(\tfrac{2\pi}{\sqrt{z}}\right)\right)\,.
\end{align}
where Erf is the error function. We then find for the resummation of $\ser$ in (\ref{SeriesInitial})
\begin{align}
\mathcal{L}^0[\mathcal{B}[\ser]](z)=\frac{2\,\pi^{3/2}}{\sqrt{z}}\,e^{\frac{4\pi^2}{z}}\,\left(1-\text{Erf}\left(\tfrac{2\pi}{\sqrt{z}}\right)\right)-1\,.\label{BorelResumS1}
\end{align}
For the appearance of the error function (or the related $\text{Erfc}(z)=1-\text{Erf}(z)$) see also \cite{Aniceto:2026pcz}.

\printbibliography
\end{document}